\documentclass[final,3p,times,twocolumn]{elsarticle}
\usepackage{graphicx}

\usepackage{multirow}
\usepackage{threeparttable}

\usepackage{hyperref}
\usepackage{amsmath}
\usepackage{amssymb}

\usepackage{subcaption}

\usepackage{caption}
\usepackage{listings}
\usepackage{xcolor} 
\usepackage{soul}
\usepackage[textsize=tiny]{todonotes}
\AtBeginDocument{%
  }

\lstdefinelanguage[myasm]{Assembler}{
  morekeywords={mov,add,sub,cmp,jne,je,jmp,call,ret,push,pop,lea,nop,int,test,dec},
  sensitive=true,
  morecomment=[l]{;},
}

\lstdefinestyle{asmstyle}{
  language=[myasm]Assembler,
  basicstyle=\ttfamily\small,
  keywordstyle=\bfseries,
  commentstyle=\itshape\color{gray},
  numbers=left,
  numberstyle=\tiny,
  stepnumber=1,
  numbersep=8pt,
  tabsize=8,
  columns=fullflexible,
  keepspaces=true,
  frame=single,
  framerule=0.4pt,
  captionpos=b
}
\lstdefinestyle{textwrap}{
  basicstyle=\ttfamily\small,
  numbers=none,
  frame=single,
  framerule=0.4pt,
  columns=fullflexible,
  keepspaces=true,
  breaklines=true,
  breakatwhitespace=false,
  showstringspaces=false
}

\begin{document}

\title{Malformer: A Multi-Modal Malware Detector Using Transformers}



\author[Tech]{Samuel Howard}
\ead{sahoward42@tntech.edu}

\author[UNE]{Kshitiz Aryal}
\ead{karyal@nebraska.edu}

\author[Ncat]{Mahmoud Abdelsalam}
\ead{mabdelsalam1@ncat.edu}

\author[Tech]{Maanak Gupta}
\ead{mgupta@tntech.edu}

\author[Tech]{Andrew Wheeler}
\ead{amwheeler43@tntech.edu}

\author[Tech]{Pradip Kunwar}
\ead{pkunwar42@tntech.edu}

\affiliation[Tech]{organization={Department of Computer Science, Tennessee Tech University},
            city={Cookeville},
            postcode={38501},
            state={TN},
            country={USA}}

\affiliation[UNE]{organization={School of Interdisciplinary Informatics, University of Nebraska at Omaha},
            city={Omaha},
            postcode={68182},
            state={NE},
            country={USA}}

\affiliation[Ncat]{organization={Department of Computer Science, North Carolina A\&T State University},
            city={Greensboro},
            postcode={27411},
            state={NC},
            country={USA}}





\begin{abstract}
Traditional malware detection systems that rely on a single representation of malware often fail to identify novel threats.
These representations of malware binaries, also known as modalities, do not provide the models with sufficient information to discriminate all samples. Additionally, individual representations introduce new failure modes with some modality extraction being dependent upon the success of disassembling.
Past works have integrated either additional modalities or more discriminative representations for classification. In this work, we present \textit{Malformer}, a quadrimodal malware detection model that incorporates text, image, graph, and audio representations of Windows executables. 
We demonstrate that multimodal transformer fusion can enhance the performance of Windows malware detectors over that of unimodal and bimodal detectors.
Malformer employs a combination of two RoBERTa encoders paired with a modified Vision Transformer for image data, WavLM for audio data, and an adaptive loss-weighting scheme to fuse modality-specific representations. Evaluated on a dataset of 201,549 binary samples, Malformer achieved 98.3\% accuracy and an F\textsubscript{1} score of 0.9833, outperforming both unimodal baselines and bimodal detectors by 4.6-17.6 percentage points. Malformer demonstrates that multimodal fusion provides a promising foundation for countering the growing scale of malware threats, equipping defenders with generalized and resilient detection capabilities.


\end{abstract}

\maketitle


%

\section{Introduction}

Modern cybersecurity practices utilize a wide variety of tools to protect digital infrastructure. Threat models for modern infrastructure are often complex, requiring a vast array of approaches to ensure organizational security policies are maintained. Of the threats that pose existential risk to security, malware is one of the most pervasive and diverse.
This can be primarily attributed to the attackers' ability to construct malicious programs and files that can abuse any number of mechanisms and channels.

Early approaches for malware detection have focused on static analysis techniques such as signature-based scanning~\cite{lo1995mcf, christodorescu2003static}.
These approaches require the derivation of unique identifiers, or ``signatures,'' from known malicious files, such as byte patterns, hash values, or instruction sequences, and storing them in a database. Such systems have the obvious limitation that they can only detect known threats, requiring manual updates for novel or uniquely obfuscated malicious samples. As a result, systems utilizing this approach are vulnerable to zero-day attacks or even minor code variations within a malware family.

Addressing some of these shortcomings, heuristic and rule-based approaches emerged~\cite{bazrafshan2013survey, dube2012malware, christodorescu2003static}. These extend detection by defining general patterns or properties indicative of malicious activity, such as self-modification, suspicious API calls, obfuscated binaries, or Virtual Machine (VM) detection. Heuristics improve generalization beyond pure signatures, though they are more prone to false positives. 
Together, these approaches established the foundation for malware detection. However, each shares a critical limitation in that they rely on handcrafted features or known malicious attributes. This human dependency exposes the need for methods more capable of generalizing to novel samples. This gap motivated the transition to machine learning (ML) based approaches.
To keep pace with adversaries and stop novel threats before they can cause serious damage, cybersecurity researchers have begun using ML algorithms to detect malicious files and classify previously unseen threats. Early applications of ML for malware detection often focused on feature engineering and classical algorithms~\cite{schultz2000data, kolter2004learning, shafiq2009pe, rieck2011automatic, arp2014drebin}. Static features such as byte n-grams, opcode frequencies, metadata fields, and imported API calls were extracted from binaries. 


As datasets grew and adversaries developed sophisticated evasion techniques, feature engineering increasingly became a bottleneck. The rise of deep learning offered the alternative of letting models learn representations directly from raw or minimally processed data, rather than hand-crafted features. Early work experimented with convolutional neural networks (CNNs) on byte sequences and images generated from binaries, and recurrent neural networks (RNNs) on API call sequences~\cite{raff2017malware, yakura2018malware, kwon2017extracting, catak2020deep}. Such approaches reduced the need for manual feature design and showed resilience to certain obfuscation strategies, though requiring substantially greater computational resources and careful tuning.
Despite their advances over traditional methods, these networks have notable limitations. CNNs, by nature of their architecture, are best for capturing local patterns such as the structure of a function within byte-level features or short instruction sequences, but struggle with long-range dependencies common to the control flow structure of a binary~\cite{alzubaidi2021review, chibuike2024convolutional}. Their fixed receptive fields and translation invariance, strengths in image domains, often fail to capture global context in code. RNNs handle sequential data more naturally, but struggle to parallelize and handle long sequences such as those found in disassembly. Both are sensitive to perturbation of their inputs, motivating a shift to more robust models.

Transformers are a neural architecture built around self-attention, a mechanism that enables the model to evaluate relationships between every pair of positions in the input sequence simultaneously~\cite{vaswani_attention_2023}.  Instead of scanning data sequentially, like RNNs, or relying on fixed receptive fields, like CNNs, transformers compute relevance globally. This allows them to be strong at modeling long-range structure, variable-length patterns, and heterogeneous inputs. For malware binaries, where important relationships can span distant parts of the byte sequence or assembly code, this ability to reason over the whole input simultaneously is a natural fit. This property of the architecture is valuable for binaries and control-flow structures where critical relationships exist not only in local areas, but also across the input. 



Previous works have demonstrated the benefits of the inclusion of multiple orthogonal views of the input binary files, up to two primary modalities or many feature vectors~\cite{arnold2000automatically, ismail2025midalf, gibert2020orthrus, schranko_de_oliveira_chimera_2020}. Each modality captures a different perspective of a binary. For example, textual representations can preserve semantic and instruction-level information, graph representations can capture structural relationships, image representations can reveal global byte-level patterns, and audio representations can encode sequential byte distributions in a transformed signal space. Relying on a single modality can therefore miss important information. We extend this prior body of work not only by our usage of transformers, but also with the combination of four modalities frequently used in malware analysis. This gives our model, Malformer, both greater discriminative power, thereby improving accuracy, and greater robustness to the failure of one or more of its modalities.

By leveraging the strengths of transformer architecture, Malformer integrates transformer backbones across modalities and fuses their outputs into a joint representation, achieving robustness to missing modalities and high detection accuracy. Unlike previous approaches, Malformer integrates four distinct orthogonal views of a program. The major contributions of this work are as follows:

\begin{itemize}
    \item We propose Malformer, a quadrimodal transformer-based architecture for Malware detection, integrating image, text, graph, and audio modalities into a single framework.
    \item We demonstrate that the combination of these heterogeneous representations yields greater performance than unimodal or bimodal baselines.
    \item We present an adaptive training strategy that balances modality contributions, preventing the overfitting of faster-converging modalities and improving overall robustness.
\end{itemize}


Rest of the paper is organized as follows. Section \ref{sec:related} summarizes the background and related works. Section \ref{sec:architecture} describes our proposed architecture and components. Section \ref{sec:experimental_setup} shows our evaluation methodology, results, and analysis of the results. Finally, Section \ref{sec:results_and_discussion} shows our results and analysis, and conclusion in Section \ref{sec:conc}.

\section{Background and Related Work}
\label{sec:related}


Malware classification has benefited from advances in deep learning across several domains, including computer vision, natural language processing, and speech processing. Many of these architectures have been adapted to operate on different representations of executable files, either individually or in combination. The following subsections review the architectures most relevant to this work and establish the foundation for the multimodal model proposed in this paper.

\subsection{Unimodal and Bimodal Classifiers}

Transformers are a machine learning architecture that utilizes an encoder and decoder to convert one data format into another. This architecture was originally proposed as an advanced method of machine translation, primarily due to its ability to find associations between related input and output data \cite{vaswani_attention_2023}.
However, since its inception, this encoder-decoder system has been adapted to several applications, including malware analysis. 

 \textbf{Vision Transformers: }
The Vision Transformer (ViT) reframes image recognition as a pure sequence modeling problem by splitting an image into fixed-size patches, linearly embedding them, and processing the resulting token sequence with a standard Transformer encoder, dispensing entirely with convolutions~\cite{dosovitskiy2020image}. By introducing a learnable class token and using positional embeddings, ViT directly applies the self-attention machinery from NLP to vision, enabling global receptive fields from the first layer. While ViT underperforms CNNs in low-data regimes due to weaker inductive biases, the paper demonstrates that, when pretrained on large-scale datasets and fine-tuned downstream, ViT matches or surpasses state-of-the-art convolutional models, highlighting the scalability of attention-based architectures and shifting the field’s focus toward data-efficient training and hybrid designs.

\textbf{ViT4Mal: }
ViT4Mal~\cite{ravi2023vit4mal} adapts Vision Transformers (ViTs) for malware classification by reformulating binaries as image-like inputs suitable for transformer-based feature extraction. In this work, executable files are converted into grayscale image representations that preserve the raw byte structure while enabling the self-attention mechanism of the vision transformer to capture both local and global relationships. Unlike CNN-based approaches such as MalConv \cite{raff2017malware}, which are limited by fixed receptive fields, the transformer layers allow ViT4Mal to model long-range dependencies across the entire executable. 
For its performance, ViT4Mal incorporates several modifications more tuned for executable-file data. First, square images are created from the input executable file's bytes, then scaled to standard dimensions. Next, the class token of the original ViT architecture is removed, and instead replaced by flattening the outputs of the last encoder layer before being used by their multilayer perceptron (MLP) decoder. 


\begin{table*}[!t]
\centering
\caption{Summary of multimodal malware detection models and their modalities.}
\vspace{-2mm}
\label{tab:multimodal_malware_comparison}
\begin{tabular}{p{1.7cm} p{1.3cm} p{7.0cm} p{3.9cm}}
\hline
\textbf{Model} & \textbf{Platform} & \textbf{Modalities Used} & \textbf{Fusion Type}\\
\hline

Orthrus~\cite{gibert2020orthrus} & Windows & 
Raw byte sequences; assembly mnemonics (opcodes) & 
Intermediate fusion\\

MIDALF~\cite{ismail2025midalf} & Windows & 
Image representations, audio spectrograms from binaries & 
Late fusion (logistic regression) \\
\hline
\end{tabular}
\end{table*}

\textbf{RoBERTa: }
RoBERTa (Robustly Optimized BERT Pretraining Approach) is a refinement of the original BERT model that demonstrates the importance of training methodology in transformer performance~\cite{liu2019roberta}. It retains the encoder-only architecture of BERT \cite{devlin2019bertpretrainingdeepbidirectional} while removing the next sentence prediction (NSP) objective, simplifying the pretraining task and improving generalization. 
Training time was also extended with larger batch sizes and longer sequences, allowing the model to capture longer-range dependencies more effectively.

All of these modifications led to improvements in benchmarks such as GLUE, RACE, and SquAD~\cite{liu2019roberta}. These results demonstrated that the underlying transformer architecture of BERT was not weak, but it was not trained at a sufficient scale. RoBERTa therefore established that longer training schedules, greater amounts of data, and simplified objectives can yield substantial gains without alteration of the underlying transformer design. This improvement in performance, particularly in the breadth it manages to capture, makes it an ideal candidate for adaptation to the task of malware classification.


\textbf{WavLM: }
Chen et al. proposed WavLM, a self-supervised framework designed to unify speech representation learning across a broad range of speech tasks~\cite{chen2022wavlm}. Previous methods in speech showed success in automatic speech recognition and phoneme classification, but struggled to generalize across tasks such as speaker diarization, separation, and verification. WavLM is designed to overcome this limitation by jointly learning masked speech prediction and denoising, enabling the model to capture both content and speaker-related characteristics more effectively. Experimental results demonstrate that WavLM achieves state-of-the-art performance on the SUPERB benchmark and a variety of other task-specific evaluations. Notably, WavLM outperforms previous models in challenging multi-speaker tasks and exhibits strong generalization across full-stack speech processing tasks. This work establishes WavLM as the first SSL model designed and demonstrated to support a comprehensive range of downstream speech applications, without requiring model-specific pre-training for individual tasks. Although designed for speech, WavLM's ability to learn robust sequential representations from raw waveforms makes it well suited for malware classification when executable files are transformed into audio, allowing the model to capture structural patterns that may not be apparent in other modalities.

\textbf{MalConv2: }
\textit{MalConv2}~\cite{raff2021classifying}, a deep learning architecture for malware detection over extremely long byte sequences, addresses the prohibitive memory and computational costs of prior raw-byte convolutional models such as MalConv~\cite{raff2017malware}. The approach introduces a fixed-memory temporal max-pooling strategy that exploits the sparsity of gradients induced by global max pooling, reducing GPU memory usage by up to $116\times$ and training time by up to $25.8\times$ while enabling end-to-end processing of executables exceeding $100$ million time steps without truncation. Building on these efficiency gains, the authors develop a novel \emph{Global Channel Gating} (GCG) mechanism that enables context-sensitive feature interactions across distant regions of an input sequence, overcoming the locality limitations of earlier architectures. Experimental results on the EMBER 2018 malware corpus demonstrate that the proposed approach improves classification performance from $91.27\%$ to $93.29\%$ accuracy and from $97.19$ to $98.04$ AUC, while simultaneously eliminating a major adversarial weakness caused by fixed input-size truncation. The work establishes a practical framework for scalable sequence classification on unprecedented input lengths and demonstrates its applicability to large-scale cybersecurity tasks.



\begin{figure*}[!t]  
    \centering
    \includegraphics[width=0.90\textwidth]{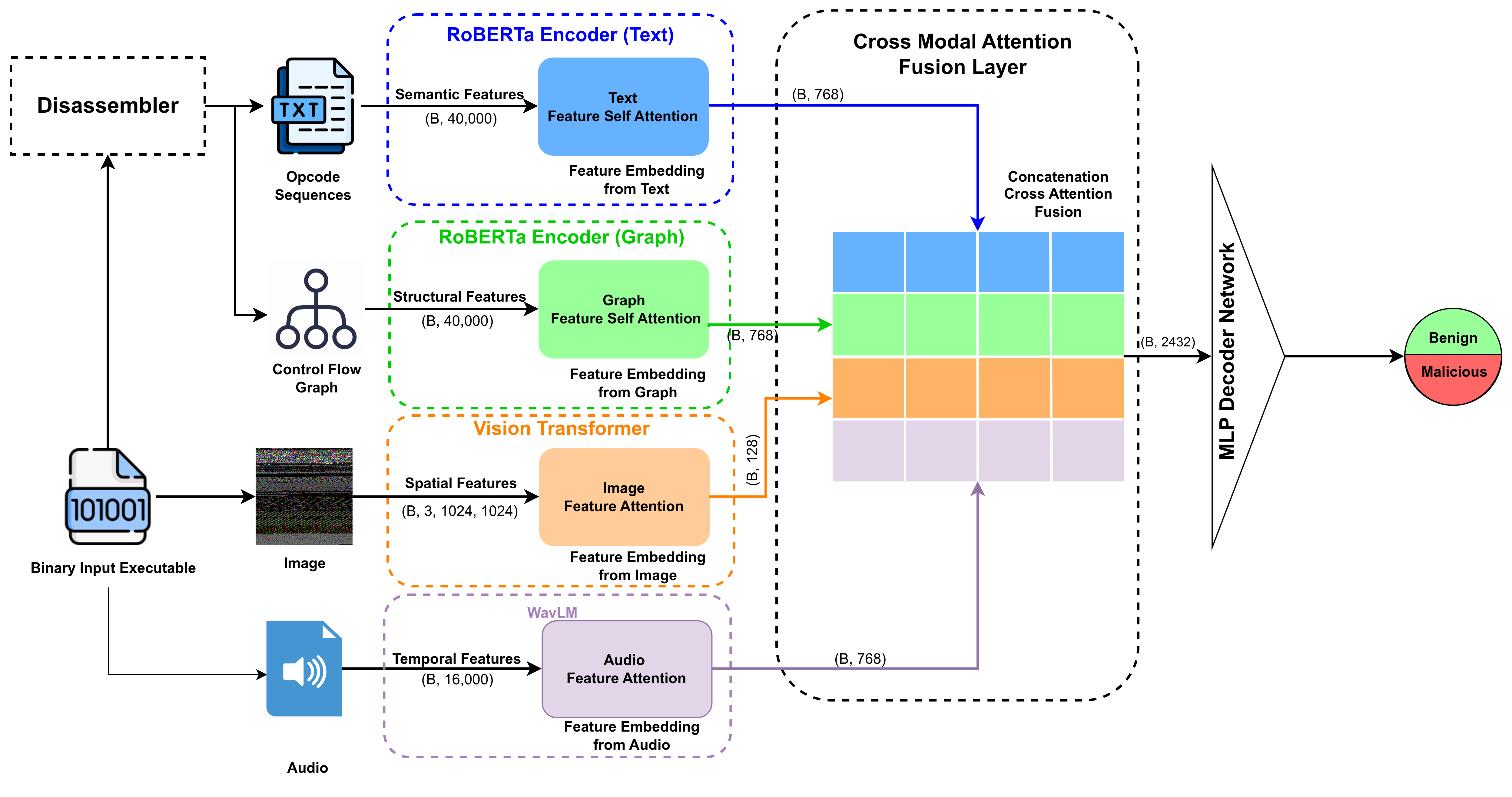}
    \vspace{-3mm}
    \caption{Malformer Architecture}
    \label{fig:architecture_diag}
\end{figure*}

\subsection{Multimodal Malware Detection}
Prior malware detectors using multiple modalities have shown promising results in their efficacy, as summarized in Table~\ref{tab:multimodal_malware_comparison}.

\subsubsection{Orthrus} Orthrus addresses the limitations of unimodal malware classifiers for Windows malware \cite{gibert2020orthrus}. The architecture combines two representations, raw byte sequences and assembly mnemonics, extracted from executable files. The authors argue that unimodal approaches miss critical complementary information and are therefore vulnerable to obfuscation strategies. Orthrus counters this by performing intermediate fusion of features learned through convolutional neural networks from both modalities, allowing for more discriminative feature spaces.
Orthrus distinguishes itself as the first end-to-end multimodal deep learning framework for malware classification of specifically Windows Portable Executable files. The architecture employs modality-specific convolutional subnetworks that extract features from byte subsequences and opcode mnemonics. These features are then fused at an intermediate stage to enable the network to leverage complementary signals across modalities. Their evaluation on the Microsoft Malware Classification Challenge dataset demonstrates that Orthrus surpasses both unimodal deep learning approaches and classical n-gram baselines, achieving higher accuracy and F\textsubscript{1}-scores across diverse malware families, including those with obfuscation.

\subsubsection{MIDALF} MIDALF is a multimodal malware detection framework that converts binaries into images and audio signals, transforming the audio signals into a spectrogram representation \cite{ismail2025midalf}. Each representation is combined using late fusion techniques. Image features are extracted using a self-supervised learning model for structural patterns, with a convolutional neural network processing audio spectrograms for temporal and frequency-based characteristics. Of the late fusion strategies tested, the authors found logistic regression to be the most effective. They achieved an accuracy of 99.7\% on the BODMAS \cite{bodmas} dataset.
The framework further demonstrated robustness against adversarial samples. Evaluated on GAN-generated samples, MIDALF maintained a detection rate of 95.1\%.  



\textit{Unlike prior unimodal and bimodal approaches, we combine a total of four orthogonal modalities trained with an adaptive loss weighting scheme. This design proved more robust for detecting malware in the case of one or more missing modalities, while strengthening in the ideal case of all modalities being present over prior work.}


\section{Proposed Malformer Architecture} 
\label{sec:architecture}
Our proposed Malformer model combines the heterogeneous sources images, control flow graphs, disassembled text, and audio within a unified multimodal transformer-based architecture, as shown in Figure~\ref{fig:architecture_diag}. The design couples a ViT for image-based representations, RoBERTa encoders for text and graph modalities, and WavLM for audio, followed by modality-specific decoders and a shared classifier. The output of each modality encoder is concatenated into a unified feature representation before being passed to the main MLP decoder network that outputs the final classification. 

Additionally, MLP decoder networks present for each modality are used for auxiliary supervision, except for the final classification. The auxiliary objectives help stabilize training by providing direct supervision to each encoder, while the joint decoder captures cross-modal dependencies. Furthermore, dropout is applied at several points to improve generalization. We allow separate dropout rates for main classifier, as well as for the image, graph, and text decoders, enabling fine-grained control over regularization across modalities.

\subsection{Vision Transformer}
We started with the PyTorch implementation of Vision Transformer (ViT) \cite{dosovitskiy2020image} architecture and hyperparameters consistent with the ViT-Base variant, a patch size of 32, an image size of 2048, and a single classification class. 
We subsequently reduced the image size to 1024 because larger resolutions increased computational cost and did not improve validation performance.

Following unsatisfactory initial results on our preliminary small dataset, we conducted a wide grid search of the parameter space by perturbing all model parameters to determine the optimal setting for the ViT. These preliminary experiments indicated that the inclusion of the \verb|.text| section of the executable played a disproportionately important role in the classification performance, with the large image resolution ultimately being detrimental. Further analysis suggested that the MLP decoder had a greater impact on performance than the transformer backbone. However, overall results remained suboptimal, prompting a deeper review of the literature. 

In our review, we identified other works that utilize ViTs for malware classification, similar to our initial approach. 
We adapted our implementation based on ViT4Mal\cite{ravi2023vit4mal}, which reported architectural modifications tailored to malware images.
As their code was not publicly available, we adapted our ViT implementation based on the architectural and methodological details provided in their paper. 
We primarily made two structural changes to our ViT implementation. Firstly, we removed the class token used by the standard ViT architecture, instead aggregating patch embeddings via mean pooling before classification. This method ensures that each patch contributes equally, preventing over-reliance on any single token and yielding a well-balanced representation. Secondly, we reduced the depth of the transformer component of the ViT incrementally from 12 to 8 layers with 4 attention heads and a hidden dimension of 128. These architectural changes improved performance, suggesting that a more compact ViT configuration is better suited for this task. 



\begin{figure}[!t]  
    \centering
    \includegraphics[width=0.45\textwidth]{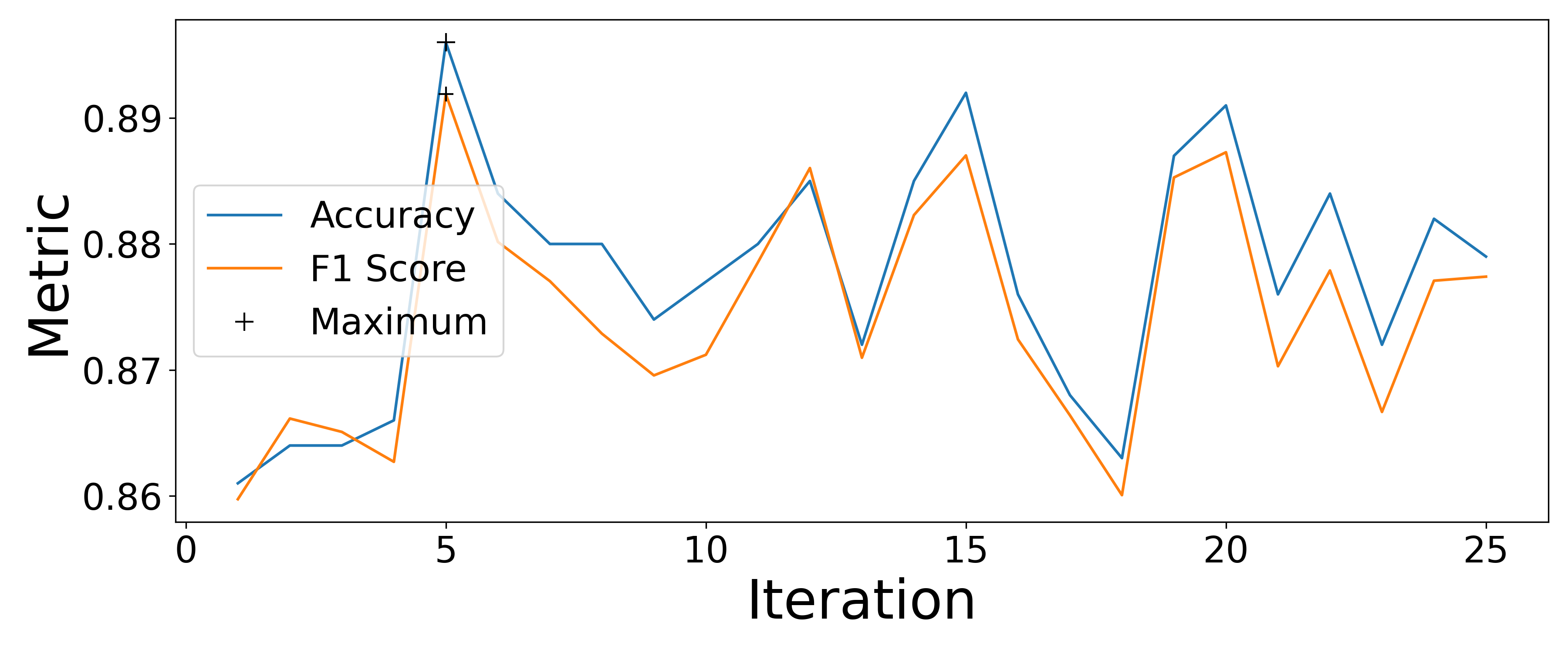}~ ~ \,
    \caption{Plot of square-patched ViT Bayesian Optimization results}
    \label{fig:bayes_plot}
\end{figure}

Using the compact ViT configuration as the starting point, we applied Bayesian optimization to tune the layer count, MLP dimensions, and decoder dropout.
We specifically chose the layer count, encoder and decoder MLP dimensions, and decoder dropout. This approach models the relationship between our chosen set of hyperparameters and performance using a Gaussian Process, providing both mean predictions and uncertainty estimates. After each iteration, a set of candidate configurations is generated and evaluated using expected improvement as the acquisition function, balancing exploration of uncertain regions and exploitation of known high-performing areas. We ran this with 5 burn-in steps and 20 optimization iterations with 200 sample candidates. The results of our optimization run can be found in Figure \ref{fig:bayes_plot}.

In addition to the standard square patch formulation, we implemented an alternative linear patching strategy that preserves the token dimensionality and downstream architecture while altering the input's spatial decomposition. Rather than extracting 2D patches via a convolutional projection, the input image is first flattened in raster order and partitioned into contiguous sequences of length $p^2$, where $p$ is the patch size. Each resulting segment is then linearly projected into the embedding space. This alternative patching strategy changes the spatial decomposition of the input and introduces a different inductive bias that may better reflect the linear byte structure of the binary.

\begin{figure}[!t]  
    \centering
    \includegraphics[width=0.45\textwidth]{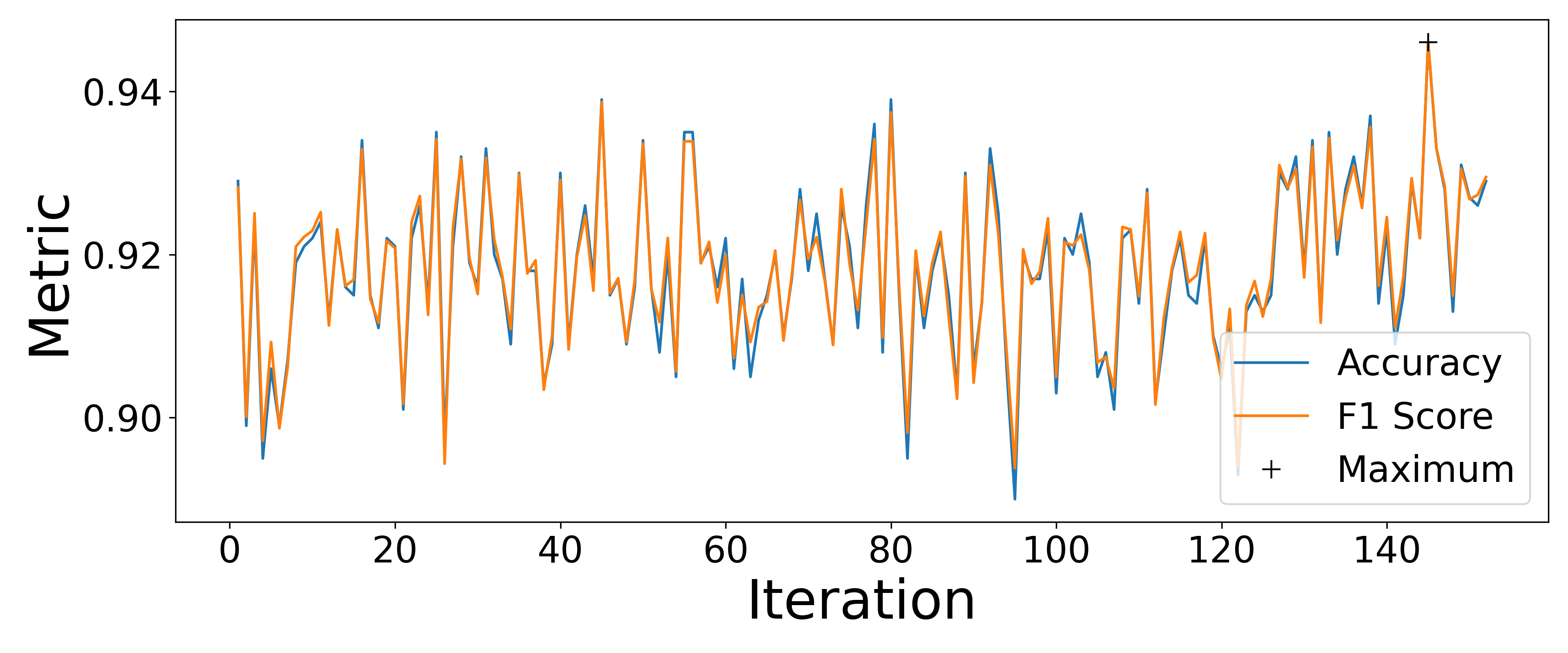}~ ~ \,
    \caption{Plot of linear-patched ViT Bayesian Optimization results}
    \label{fig:bayes_plot_lvit}
\end{figure}

We again ran Bayesian Optimization with 5 burn-in steps and 150 optimization iterations with 200 sample candidates for this linear model. Bayesian optimization did not improve on the initial hyperparameter configuration. The results of our optimization run can be found in Figure \ref{fig:bayes_plot_lvit}.

\subsection{RoBERTa}

Of all the architectures tested, the Huggingface transformers\cite{wolf2019huggingface} implementation of RoBERTa was the most performant. Before choosing RoBERTa for graph and text, we investigated representation learning approaches such as node2vec \cite{grover2016node2vec} and Asm2vec \cite{ding2019asm2vec}. However, these methods proved prohibitively computationally expensive and memory-intensive to train at the scale of our dataset, and their performance was barely greater than a random guess. This poor performance is likely due to shortcuts taken to address these issues. By contrast, RoBERTa consistently performed well for both graph and text modalities, leveraging its capacity to model long-range dependencies. Its empirical reliability, coupled with practical efficiency, made RoBERTa the most effective choice for these modalities.

For hyperparameters, we chose Ro\-BERTa\textsubscript{\, BASE}\cite{liu2019roberta} with a sequence length of 1024, Binary Cross Entropy Loss, and the Adam optimizer with a learning rate of 0.001 for testing. As is common practice, we used pre-trained weights for the two RoBERTa-based modalities. Despite the significantly different intended domain application, the models performed well. We initially did not use pre-trained weights for RoBERTa, but using pre-trained weights yielded a mild improvement in accuracy.

For the graph modality, RoBERTa's default byte-level tokenizer could be applied directly to serialized representations without modification. In contrast, text (assembly code) required the creation of a new tokenizer to allow the input data to be properly encoded. To address this, we created a new Byte-Level BPE tokenizer using the Hugging Face \texttt{tokenizers} library, trained on the totality of the text corpus. The tokenizer was configured with a vocabulary size of 50,265 tokens (the default RoBERTa setting), a minimum token frequency threshold of 2, and the standard RoBERTa special tokens \texttt{<s>}, \texttt{<pad>}, \texttt{</s>}, \texttt{<unk>}, and \texttt{<mask>}.


Tuning of the hyperparameters focused on the MLP decoder of each model. During testing, dropout was manually varied from $0.0$ to $0.5$ in steps of $0.1$. The best results for each modality, $0.5$ and $0.3$ for graph and text, respectively, were used as the dropout parameter for each modality's decoder MLP in the quadrimodal architecture of Malformer. This difference suggests that the graph representation benefited from stronger regularization than the text representation, likely due to its denser structure.

\subsection{WavLM}

Similar to the RoBERTa-based models, we utilized the implementation of WavLM~\cite{chen2022wavlm} from the Huggingface transformers library\cite{wolf2019huggingface}. Although our focus was not on benchmarking multiple audio models, WavLM met two important practical requirements in that it produced compatible embeddings that could be integrated into our existing cross-modal attention fusion layer without modification, and it performed on par with RoBERTa in early experiments with minimal additional compute and memory requirements. WavLM's combination of low computational overhead and high performance proved to be a natural complement to the ViT in this architecture.

Like RoBERTa, we utilized pretrained weights for WavLM. Each binary was converted into a 10-second, single-channel PCM waveform at a sampling rate of 16~kHz, as described in Section~\ref{feature_extraction}. The raw waveform was fed directly to WavLM, which outputs a sequence of 768-dimensional hidden states corresponding to local temporal windows of the signal. To obtain a fixed-length embedding suitable for classification, we applied mean pooling across the temporal dimension, resulting in a single 768-dimensional vector per sample. This embedding was then passed to the appropriate MLP(s) for decoding into a classification of a particular sample.

\subsection{Cross-Modal Attention Fusion}

A central challenge in constructing a multi-modal model is determining an effective strategy for integrating modality-specific components. In their survey, Xu et al. \cite{xu2023multimodal} provide a structured taxonomy of architectural designs for combining a plurality of data modalities within machine learning models. Among these methods, Early Summation, Early Concatenation, and Multi-Head Self-Attention (MHSA) represent three key strategies for fusing modality representations. 

In our work, Early Summation was combined with MHSA such that the resulting logits of the summation were used as the query, key, and value of MHSA. 
Early Concatenation can be defined as follows. Given logits $\text{text} \in \mathbb{R}^{\{n \times a\}}$, $\text{image} \in \mathbb{R}^{\{n \times b\}}$, $\text{graph} \in \mathbb{R}^{\{n \times c\}}$, $\text{audio} \in \mathbb{R}^{\{n \times d\}}$. Then
\[
C = [\text{text} | \text{image} | \text{graph} | \text{audio}] \in \mathbb{R}^{\{n \times (a + b + c + d)\}}
\]
where $[\text{text} | \text{image} | \text{graph} | \text{audio}]$ denotes the column-wise concatenation of the matrices. The matrix $C$ can then passed through the encoder(s). Early Concatenation was likewise combined with MHSA such that the resulting logits of the concatenation were used as the query, key, and value of MHSA. 
Of the three we tested, Early Concatenation without MHSA performed the best in terms of accuracy and F\textsubscript{1} score, and was therefore used for Malformer.

\subsection{Modality Combination} \label{malformer_method}

The proposed architecture integrates four modality-specific encoders, including two independent Ro\-BERTa\textsubscript{BASE} \cite{liu2019roberta} backbones for the text and graph modalities, a modified Vision Transformer (ViT) \cite{ravi2023vit4mal} for the image modality, and WavLM for the audio modality. Following individual encoding, the modality representations are concatenated by the Early Concatenation method and passed to the multimodal decoder MLP. Each modality also has an additional classification head exclusively for training that does not contribute to the multimodal prediction.

Training utilized a joint loss function joining modality-specific prediction losses with an ensemble head loss, a process called auxiliary supervision. The total loss at epoch $t$ is expressed as  
\[
\mathcal{L}_t = \lambda_E \, \mathcal{L}_E + \sum_{m=1}^M \lambda_m^{(t)} \, \mathcal{L}_m
\]
where $\mathcal{L}_E$ denotes the multimodal head loss, $\mathcal{L}_m$ is the loss for modality $m$, $\lambda_E$ is the fixed multimodal loss weight, and $\lambda_m^{(t)}$ is the dynamically updated weight for modality $m$ at epoch $t$. In this work, $\lambda_E = 0.2$ and $M=4$ (text, image, graph, audio), with the image modality initialized with a larger weight of 0.65 and the audio modality initialized to 0.15 to compensate for its slower convergence. We chose to initialize the image modality with such a high weighting after preliminary experiments showed overfitting of the RoBERTa-based modalities before the convergence of the image modality.

Modality weights $\lambda_m^{(t)}$ evolve over training using an allocation proportional to the error with exponential moving average (EMA) smoothing. Let $a_m^{(t)}$ be the accuracy of modality $m$ at epoch $t$, and define the error $e_m^{(t)} = 1 - a_m^{(t)}$. The non-normalized modality weights are  
\[
\tilde{w}_m^{(t)} =
\begin{cases}
\frac{e_m^{(t)}}{\sum_{j=1}^M e_j^{(t)}}, & \text{if } \sum_{j=1}^M e_j^{(t)} > 0 \\
\frac{1}{M}, & \text{otherwise}
\end{cases}
\]
These are scaled to sum to $(1 - \lambda_E)$, then smoothed with EMA according to  
\[
\hat{w}_m^{(t)} = \alpha \, w_m^{(t-1)} + (1 - \alpha) \, \tilde{w}_m^{(t)}
\]
where $\alpha \in [0, 1]$ is the EMA smoothing factor (here, $\alpha = 0.99$). The smoothed weights are finally renormalized to sum to $(1 - \lambda_E)$:
\[
\lambda_m^{(t)} = \frac{\hat{w}_m^{(t)}}{\sum_{j=1}^M \hat{w}_j^{(t)}} \cdot (1 - \lambda_E)
\]

For multimodal training, the optimal hyperparameters for each modality, as determined from unimodal experimental results in Section~\ref{unimodal-results}, were retained in the larger model. This was done to ensure the optimal performance of each modality to promote the performance of the larger model. The larger architecture introduced an additional MLP decoder with its associated hyperparameters. A hidden size of 3,072 was used, and dropout was manually tested from 0.0 to 0.5 with steps of 0.1.

\subsection{Multilayer Perceptron Decoders}
For each modality, the output of the encoder backbone is projected to a scalar value with a lightweight multilayer perceptron (MLP) decoder. The decoder accepts a fixed-length feature vector, corresponding to a modality-specific encoder or the overall model. The MLP consists of a fully connected layer mapping the input to a hidden representation, followed by a rectified linear unit (ReLU) activation function, and dropout regularization. The final, fully-connected layer reduces this hidden representation to a single scalar value. This value represents the raw prediction score for a given input sample. This structure was chosen for its simplicity, while retaining sufficient capacity to transform the encoder’s high-dimensional feature representation into a prediction. 

\section{Experimental Setup} \label{sec:experimental_setup}


All experiments were implemented in PyTorch version 2.5.1 along with compatible helper libraries like NumPy \cite{paszke2019pytorch, harris2020array}. All of our models that we trained utilized PyTorch's Automatic Mixed Precision (AMP) package for acceleration. Feature extraction, training, and evaluation were conducted on two separate environments for isolation purposes. Firstly, a server running Ubuntu 22.04.5 was used for feature extraction and experiments involving smaller model architectures. Next, a shared high-performance computing (HPC) cluster running Rocky Linux 8.9 was utilized to train our larger architecture and some of the independent modalities. This setup was required to utilize the higher-performance GPUs available on the HPC cluster and to minimize the potential risk to other users. 

The Ubuntu server was equipped with an AMD Ryzen 5700x CPU, an NVIDIA RTX 3060 12 GB GPU, and 48 GB of system memory. Next, each node of the HPC cluster used had dual AMD EPYC 7713 CPUs, dual NVIDIA A100 PCIe 40 GB GPUs, and 512 GB of system memory. Of this, only one CPU and up to two GPUs were used in parallel for training, with one GPU being used for evaluating model performance. Both environments utilized CUDA-enabled GPUs of the same microarchitecture for model training acceleration. This also had the added benefit of ensuring numerical consistency between GPUs of matching compute capability and allowing for quicker iterative development.

\subsection{Dataset} \label{dataset_collection}

To evaluate each architecture, we utilized two datasets: one internal dataset, which is intentionally difficult, and a second external dataset for reproducibility.

\begin{table*}[!t]
\centering
\caption{Comparison of Smaller and Larger Datasets}
\vspace{-2mm}
\label{tab:dataset_comparison}
\begin{tabular}{lcccccc}
\hline
\multicolumn{7}{c}{\small{Small dataset}} \\
\hline
Category  & Original Files & Valid PE Files & Full Feature Extraction & Assembly & Graph & Image/Audio \\
\hline
Benign    & 33,246 & 7,079  & 4,693 & 5,504  & 6,501  & 7,079 \\
Malicious & 12,634 & 12,597 & 4,177 & 6,377  & 8,791  & 12,597 \\
All       & 45,880 & 19,676 & \textbf{8,870} & 11,881 & 15,292 & 19,676 \\
\hline
\multicolumn{7}{c}{\small{Large dataset}} \\
\hline
Category  & Original Files & Valid PE Files & Full Feature Extraction & Assembly & Graph   & Image/Audio \\
\hline
Benign    & 86,812  & 86,812  & 59,918  & 82,446  & 60,680  & 86,812 \\
Malicious & 114,737 & 114,737 & 81,924  & 86,744  & 98,124  & 114,737 \\
All       & 201,549 & \textbf{201,549} & 141,842 & 169,190 & 158,804 & 201,549 \\
\hline
\end{tabular}
\end{table*}

\textbf{Internal Dataset:}  To build our initial dataset, we collected samples from multiple sources to capture a wide range of both malicious and benign Windows executables. Benign samples were primarily obtained from a clean installation of Microsoft Windows 10, representing native system executables and commonly included software. Additional benign files were sourced from publicly available software archives on \texttt{archive.org}. All benign candidates were scanned using both Windows Defender and ClamAV to reduce the likelihood of polluting the benign set with malicious files. Furthermore, malicious samples were obtained from the VirusTotal repository. These samples represent an aggregate of malicious executables from a wide variety of malware families.
All collected samples underwent a validation process to ensure that they comply with the Portable Executable (PE) file format. Files determined to be invalid PE files were excluded from the dataset. Additionally, for this dataset, we only use the samples for which all feature extraction succeeded for a ``perfect'' case for the models tested.

The final dataset was split into disjoint training and evaluation sets. The evaluation set contained a total of 1,000 samples with an even split of 500 benign and 500 malicious samples. The training set contained a total of 8,870 samples with a split of 4,177 malicious files and 4693 benign samples. This split was chosen to provide both a robust training and validation set for balanced performance measurement throughout the acquired samples. Furthermore, the even split of the evaluation set allows better intuition regarding the performance of evaluated models, as a random guess would yield an approximate accuracy of 50\%. The training set, by contrast, has a mild benign bias, with benign samples representing approximately 53\% of the total training set.
We used this dataset for rapid iteration of the parameter space of each of our models, as well as initial evaluation metrics. Furthermore, this dataset is more relevant for the replication of the results of the CNN-based models tested, as they were trained on a relatively smaller dataset. This dataset is referred to as ``Small dataset'' both in analysis and when evaluation metrics are reported.

\textbf{External Dataset Integration: } In addition to the dataset we constructed, we incorporated the publicly available PE Malware Machine Learning Dataset provided by Practical Security Analytics \cite{Lester_PEMalwareDataset}. This dataset consists of raw PE files categorized as either benign or malicious. It contains a substantially larger number of samples than our initial dataset, with over 201,549 files in total, including 114,737 malicious and 86,812 benign executables.
We integrate this dataset for two primary reasons. First, incorporating a significantly larger and more diverse corpus of samples allows us to evaluate the generalization performance of our models beyond the scope of our curated dataset. This is particularly important for the transformer-based models we integrate into our architecture, such as RoBERTa. Second, using a publicly available dataset supports reproducibility, enabling others to replicate and extend our findings.

Using this dataset, we followed a similar methodology in terms of data cleaning. Our validation process did not exclude any samples, so the entire set of 201,549 samples was used. Instead of a static 1,000 samples, we extracted 10\% of the samples to create an evaluation set. Training and evaluation sets are disjoint, with a total of 181,395 samples for training and 20,154 samples for evaluation. As with the internal dataset, the evaluation set is an even split such that a random guess would yield an approximate accuracy of 50\%. Unlike our dataset, this leaves a slight malicious bias in training, with malicious samples representing approximately 57.7\% of the training set. Additionally, this dataset includes samples for which feature extraction was not successful.

We did not mix this dataset with the small dataset, primarily for reproducibility. Instead, the Practical Security Analytics dataset was treated as a separate benchmark to measure how well our models scale to substantially larger, more diverse datasets. This dataset is hereafter referred to as ``Large dataset'' both in analysis and when evaluation metrics are reported.

\textbf{Dataset Comparison: } 
For both datasets used, we evaluated the success of the feature extraction pipeline, as summarized in Table \ref{tab:dataset_comparison}. Both image and audio extraction always succeed as they purely rely on there being data to be read. By contrast, both assembly and graph extraction require successful disassembly. Malware authors can use a variety of techniques to obfuscate their code from this part of the pipeline, leading to failures therein.


\subsection{Feature Extraction} \label{feature_extraction}


For inference on a particular executable file, the modality features must be extracted from the source. In the case of Malformer, we extract the x86-64 assembly code from a Windows Portable Executable (PE) format file, generate a call graph, and transform the bytes of the file into an image and an audio file. 

\textbf{Assembly Code: }
To extract assembly code from a given executable file, we utilize the Radare2 reverse engineering framework. With this, we extract code from the \verb|.text| section of the file. We next normalize the features of the extracted code, such as registers and immediate operands, to reduce the token space required for the encoder model. An example of this output can be found in Figure~\ref{fig:asm-loop}.

\begin{figure}[!t]
\centering
\begin{minipage}{0.45\columnwidth}
\begin{lstlisting}[style=asmstyle, linewidth=\linewidth]
je val
mov reg16, val
call val
call val
dec reg16
mov reg16, val
\end{lstlisting}
\end{minipage}%
\hfill
\begin{minipage}{0.45\columnwidth}
\begin{lstlisting}[style=asmstyle, linewidth=\linewidth, firstnumber=7]
mov reg16, reg16
mov reg16, reg16
call val
dec reg16
mov reg16, val
cmp reg16, val
\end{lstlisting}
\end{minipage}
\caption{Output of normalized disassembler (truncated)}
\label{fig:asm-loop}
\end{figure}

\textbf{Control Flow Graph: }
Control flow graphs (CFGs) were extracted from the target PE files using the \emph{Radare2} disassembly framework. More specifically, the Python bindings made available in the \emph{r2pipe} Python package were used. Analysis yielded JSON data containing all of the calls made by each function.
To more easily handle the data, the \emph{NetworkX} Python package was used. 
The raw JSON data created by r2pipe was preprocessed, and the caller-callee relations were captured in a directed graph (digraph) data structure.

Once a digraph was constructed, one final analysis was performed to obtain the final graph-based representation of the executable. This involved traversing the graph and recording the function addresses as they were encountered in the traversal. Breadth-first traversal was performed, with the depth of the call directly correlating with how late it was recorded as being ``visited.'' To prevent loops from causing infinite recursion, duplicate appearances of a function were omitted from the final set. This set was then saved into a text file as a list of function addresses. An example of these extracted calls can be found in Figure \ref{fig:function_calls}.

\begin{figure}[!t]
\centering
\begin{minipage}{0.90\columnwidth}
\begin{lstlisting}[style=textwrap, linewidth=\linewidth]
'entry0' 'sym.__security_init_cookie' 'sym.imp.KERNEL32.dll_GetSystemTimeAsFileTime' 'sym.imp.KERNEL32.dll_GetCurrentProcessId' 'sym.imp.KERNEL32.dll_GetCurrentThreadId' 'sym.imp.KERNEL32.dll_GetTickCount'
\end{lstlisting}
\end{minipage}
\caption{Function calls extracted from a binary (truncated)}
\label{fig:function_calls}
\end{figure}

\textbf{RGB Image: } Generation of both grayscale and color images from executable files has already been shown to be a feasible endeavor. We opted to use red-green-blue (RGB) image representations of executable files. Triplets of bytes were extracted from the executable to form the red, green, and blue color channels.

\begin{figure}[t]
  \centering
  \begin{minipage}{0.45\linewidth}
    \centering
    \begin{subfigure}{\linewidth}
      \centering
      \includegraphics[width=\linewidth]{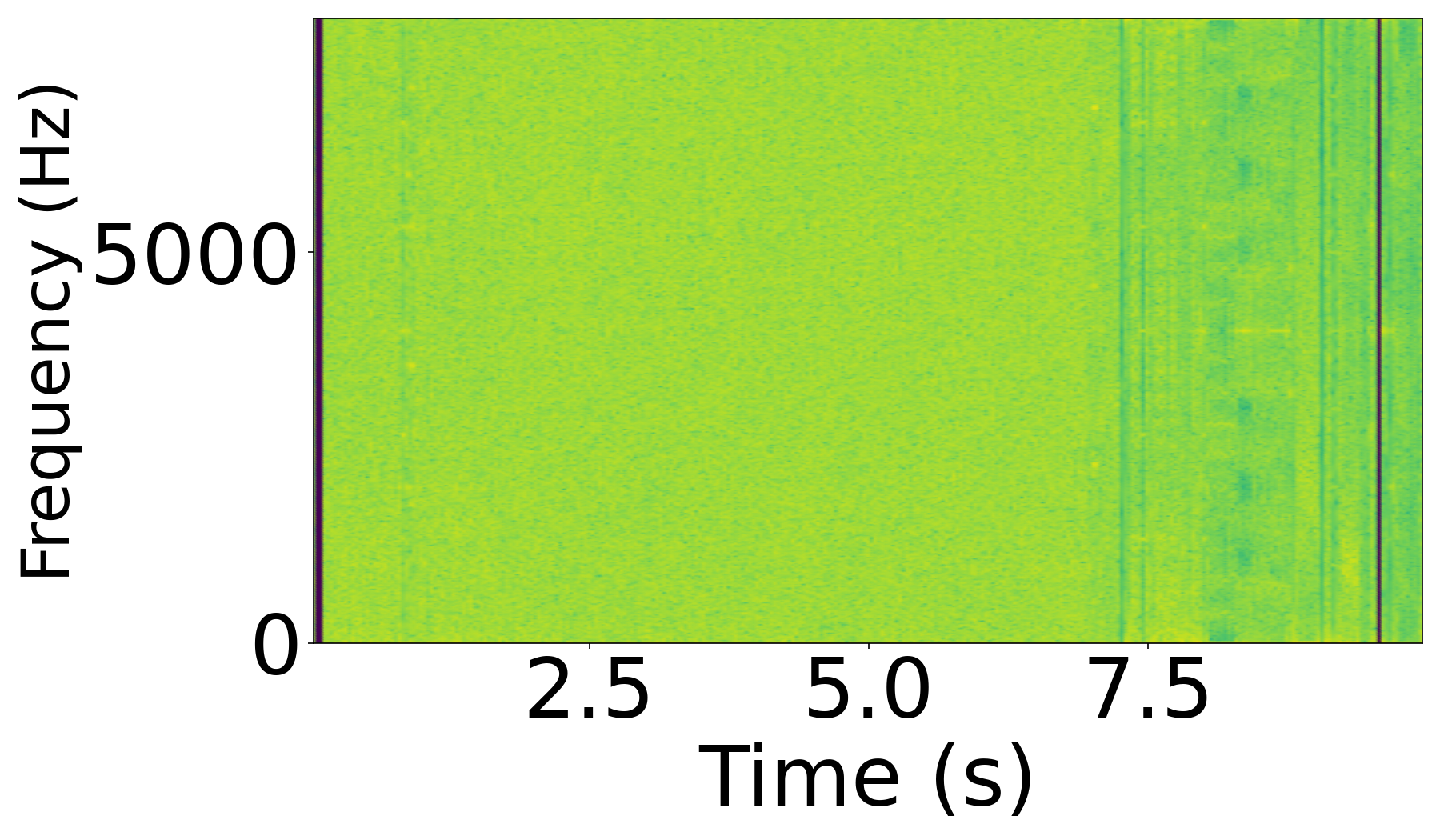}
      \caption{Benign spectrogram}
      \label{fig:audio_spec}
    \end{subfigure}

    \vspace{0.5em}

    \begin{subfigure}{\linewidth}
      \centering
      \includegraphics[width=\linewidth]{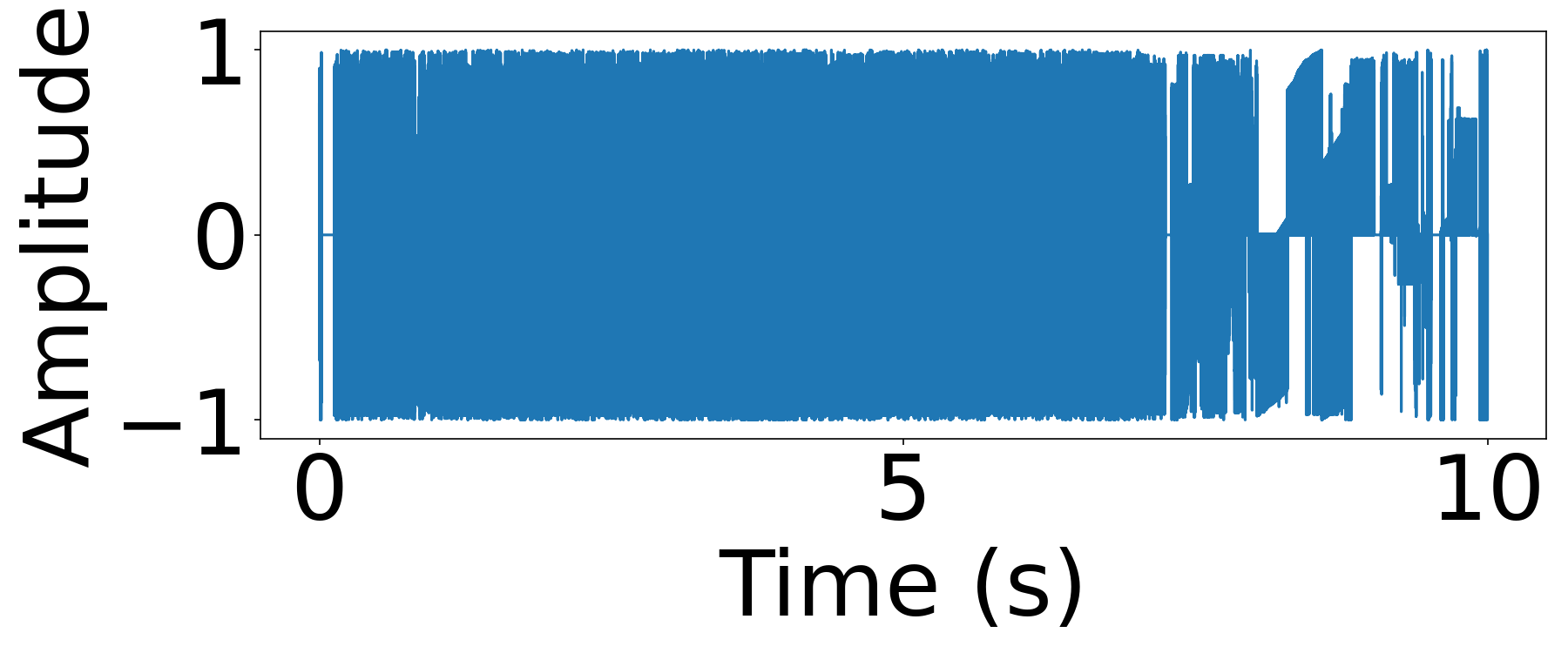}
      \caption{Benign waveform}
      \label{fig:audio_wav}
    \end{subfigure}
  \end{minipage}
  \hfill
  \begin{minipage}{0.45\linewidth}
    \centering
    \begin{subfigure}{\linewidth}
      \centering
      \includegraphics[width=0.5\textwidth]{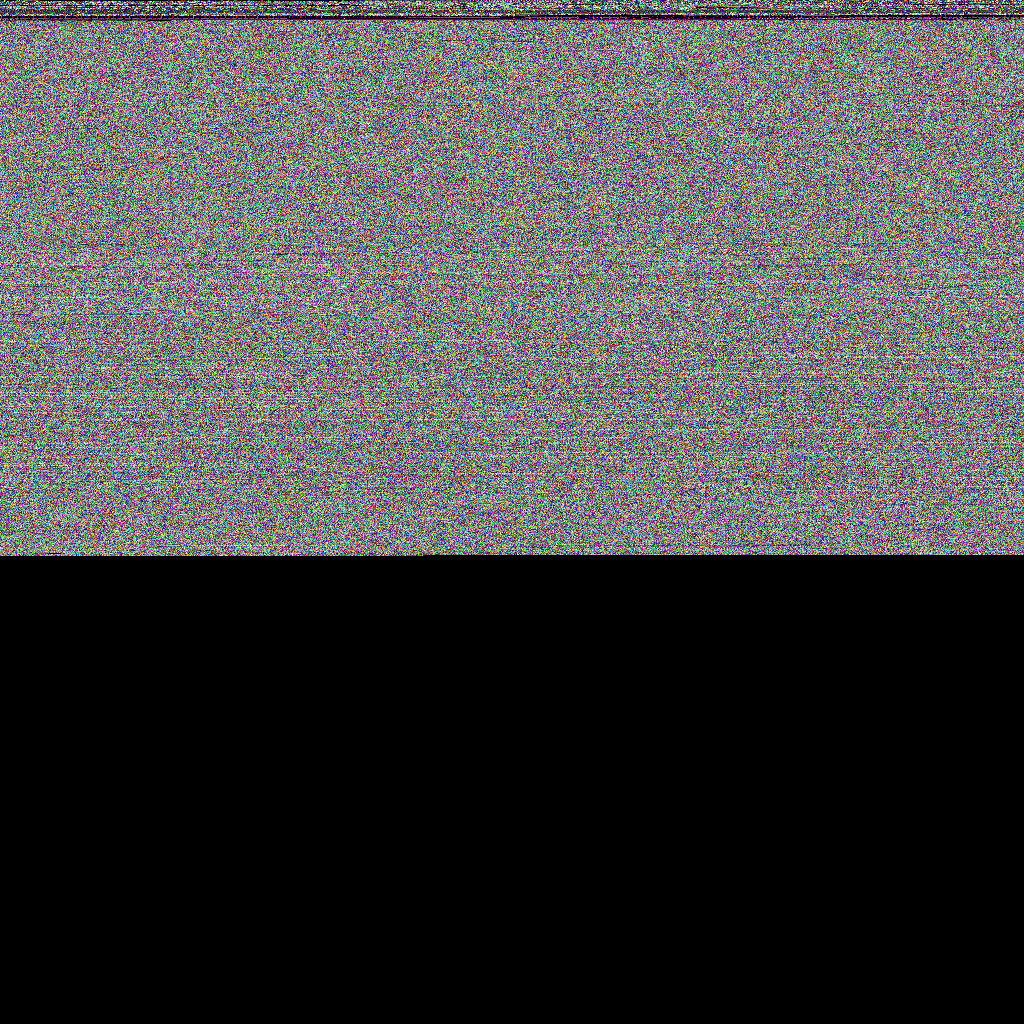}
      \caption{Image with padding}
      \label{fig:rgb_img}
    \end{subfigure}
  \end{minipage}

  \caption{Example of image and audio representations}
  \label{fig:audio_stack}
  \vspace{-5mm}
\end{figure}

A maximum of three mebibytes ($3 \times 2^{20}$) were extracted from each input file to create the corresponding $1024 \times 1024$ color image. Executable files that exceeded the three mebibyte limit were truncated, and all subsequent bytes were ignored. Conversely, executable files smaller than the defined size were padded with null bytes. An example of this can be seen in Figure \ref{fig:rgb_img}.


\textbf{Audio: }
To extract an audio-based representation of the binary, we interpret the raw file bytes as 16-bit signed integer audio samples. A fixed-length segment of 10 seconds is generated at a sampling rate of 16 kHz, resulting in 160,000 samples per file. Executables smaller than the required byte length are padded with zeros, while larger files are truncated. The resulting waveform is written to a standard WAV file with a single audio channel and 16-bit PCM encoding. An example of this can be found in Figure \ref{fig:audio_spec} and Figure \ref{fig:audio_wav}.

\subsection{Evaluation Metrics}


Our primary evaluation metrics are accuracy and F1 score. We also report the false positive rate (FPR) and false negative rate (FNR) for reference, though these play a secondary role in our analysis. All metrics are computed with a fixed decision threshold of 0, where predictions greater than zero are labeled malicious, and those less than or equal to zero are labeled benign.

\begin{table*}[!t]
\centering
\caption{Feature Extraction Performance Metrics}
\vspace{-3mm}
\label{tab:feature_extraction}
\begin{tabular}{lcccc}
\hline
\textbf{Feature} & \textbf{Time taken (total)} & \textbf{Time per sample} & \textbf{Samples per second} & \textbf{Limitation} \\
\hline
Image      & 32m 26s            & 0.009658s    & 103.53      & I/O throughput + CPU\\
Text (asm) & 24m 31s            & 0.007302s    & 136.93      & I/O throughput \\
Graph      & 16h 57m 9s         & 0.302804s    & 3.3024      & CPU \\
Audio      & 25m 4s             & 0.007464s    & 133.96      & I/O throughput \\
\hline
\end{tabular}
\end{table*}
\section{Results and Discussion} \label{sec:results_and_discussion}

\subsection{Feature Extraction Performance}

A key feasibility aspect of the deployment of malware classifiers is the time required for their feature extraction stage. To this end, we separated and measured each part of the pipeline while extracting the features of the entire large dataset on the Ubuntu server. Extraction of each modality utilized all 16 available threads. The results of this can be found in Table \ref{tab:feature_extraction}.

Of the four modalities, image, text, and audio extraction took nearly the same amount of time at roughly 32, 24, and 25 minutes, respectively. This is due to all three being mostly limited by the read/write throughput of the disk being utilized. Notably, of the three, the image extraction was the slowest. This is primarily due to the utilization of a PNG compression level 9 for OpenCV to save disk space. This caused all 16 threads to access the disk simultaneously, alternating between an I/O bottleneck and a CPU bottleneck. However, this could be made purely an I/O bottleneck by using the default PNG compression level.

In contrast, graph extraction was orders of magnitude slower, averaging just over 3 samples per second. This delay arose from the computational burden of CFG reconstruction and subsequent normalization, which are CPU-intensive and not amenable to trivial intra-sample parallelization. The disparity highlights a key deployment trade-off for graph-based features in that they are semantically rich but come with significantly higher preprocessing overhead. 
Still, it is practical as an example for a real-world pipeline that would more tightly couple each extraction step to increase speed by amortizing read penalties.

\subsection{Unimodal Results} \label{unimodal-results}

To establish performance baselines for each modality, we trained and evaluated each of the modality-specific encoders described in Section \ref{dataset_collection} on the evaluation set of 1,000 samples (500 malicious, 500 benign) for the small dataset and the evaluation set of 20,154 samples (10,077 malicious, 10,077 benign) for the large dataset. Specifically, the subset of the larger dataset with all features present was used for this comparison. This was done for consistency between modalities for fairness. Table~\ref{tab:unimodal_results} summarizes the results in terms of classification accuracy and F\textsubscript{1} score. Table~\ref{tab:unimodal_confusion_matrix} shows the confusion matrices for each dataset. Each modality demonstrates strong individual performance, but the graph modality yields the highest accuracy and F\textsubscript{1} score on the small dataset, whilst the image modality with linear patches yields the highest accuracy and F\textsubscript{1} score on the large dataset. For the graph, this suggests a greater discriminative power in the features of the control flow graph compared to text representations. For the image modality, relative to the square-patched baseline, this demonstrates the benefits of the structural bias presented by linear patching.

The image modality with square patches is the least performant modality at 89.2\% and 90.9\% accuracy for the Small and Large datasets, respectively. The image modality introduces a 2D spatial component to a dense input, rather than being purely linear. This can be useful for identifying local features of the input binary, but it yields relatively worse performance compared to the other modalities. However, input binary can always be represented as an image, whereas CFGs and text data may fail to be extracted. Thus, while the performance of the image modality is worse, images are more robust as an input.

The audio modality also performs strongly. It surpasses both image and text-based methods on the small dataset and matches the text modality on the large dataset. The WavLM encoder can extract useful temporal-shape and frequency-based signals from the binary-as-waveform representation. However, the modality lacks explicit binary semantics and may be sensitive to preprocessing design choices, which may constrain its generalizability. Even so, the strong performance with relatively lightweight preprocessing suggests that byte-derived audio signals are a promising and underexplored representation for malware classification.


Finally, the text modality performs between the best-performing graph and image variants and the weaker square-patch image baseline. It captures instruction-level information, but unlike CFGs it lacks the structural context of the original binary. Overall, the unimodal results show that each representation provides different and complementary information, and that performance varies with both dataset scale and the choice of encoding.

\begin{table*}[!t]
\centering
\caption{Unimodal performance comparison}
\vspace{-3mm}
\label{tab:unimodal_results}
\begin{tabular}{lllll}
\hline
\multirow{2}{*}{\textbf{Modality}} & \multicolumn{2}{c}{{\small\textbf{Accuracy (\%)}}} & \multicolumn{2}{c}{\small{\textbf{F$_1$ Score}}} \\
 & \scriptsize{Small dataset} & \scriptsize{Large dataset} & \scriptsize{Small dataset} & \scriptsize{Large dataset} \\
\hline
Image (Square) & 89.2 & 90.9 & 0.8893 & 0.9082 \\
Image (Linear) & 92.0 & \textbf{95.4} & 0.9192 & \textbf{0.9540} \\
Graph & \textbf{94.2} & 94.6 & \textbf{0.9447} & 0.9473 \\
Text & 91.5 & 93.7 & 0.9162 & 0.9375 \\
Audio & 93.0 & 92.9 & 0.9280 & 0.9321 \\
\hline
\end{tabular}
\end{table*}

\begin{table*}[!t]
\centering
\caption{Unimodal Confusion Matrices for Both Datasets}
\vspace{-3mm}
\label{tab:unimodal_confusion_matrix}
\begin{tabular}{lcccc|cccc}
\hline
 & \multicolumn{4}{c}{\small{Small Dataset}} & \multicolumn{4}{c}{\small{Large Dataset}} \\
\textbf{Model} & \footnotesize{\textbf{TP}} & \footnotesize{\textbf{TN}} & \footnotesize{\textbf{FP}} & \footnotesize{\textbf{FN}} & \footnotesize{\textbf{TP}} & \footnotesize{\textbf{TN}} & \footnotesize{\textbf{FP}} & \footnotesize{\textbf{FN}} \\
\hline
Text & 459 & 457 & 43 & 41 & 6642 & 6656 & 436 & 450 \\
Graph & 478 & 466 & 34 & 22 & 6664 & 6779 & 313 & 428 \\
Image (Square) & 434 & 458 & 42 & 66 & 6354 & 6546 & 546 & 738 \\
Image (Linear) & 455 & 465 & 35 & 45 & 6789 & 6740 & 352 & 303 \\
Audio & 451 & 479 & 21 & 49 & 6860 & 6324 & 768 & 232 \\
\hline
\end{tabular}
\end{table*}

\subsection{Multimodal Results}

\begin{table*}[!t]  
\centering
\caption{Multimodal Performance Comparison}
\vspace{-3mm}
\label{tab:multimodal_results}
\begin{tabular}{lcllll}
\hline
\multirow{2}{*}{\textbf{Model}} & \multirow{2}{*}{\textbf{Modalities}} & \multicolumn{2}{c}{\small{\textbf{Accuracy (\%)}}} & \multicolumn{2}{c}{\small{\textbf{F$_1$ Score}}} \\
& & \scriptsize{Small dataset} & \scriptsize{Large dataset} & \scriptsize{Small dataset} & \scriptsize{Large dataset} \\
\hline
MalConv2 & byte sequence & 95.4 & 97.7 & 0.9533 & 0.9766 \\
Orthrus & image, text  & 91.3 & 81.6 & 0.9093 & 0.8165 \\
MIDALF & image, audio & 91.4 & 91.4 & 0.9128 & 0.9204 \\
Malformer with modality voting (ours) & image, graph, text, audio & 81.2 & 97.6 & 0.8406 & 0.9758 \\
Malformer (ours) & image, graph, text, audio & \textbf{97.1} & \textbf{98.3} & \textbf{0.9758} & \textbf{0.9833} \\
\hline
\end{tabular}
\end{table*}

\subsubsection{Classification performance}

We evaluated multiple multimodal architectures, including two bimodal baselines, Orthrus and MIDALF, and the proposed quadrimodal Malformer. Orthrus combined image and text modalities, MIDALF combined image and audio, and Malformer integrated image, graph, text, and audio modalities. For Orthrus and MIDALF, we used their code, modified only to align with the file structure of our dataset and to add extra logging output. For the small dataset, we used the set of samples with all features present. For the larger dataset, we used the entire dataset, including samples with missing features. We included samples without all features present to evaluate the performance of the model with the inclusion of cases in which the feature extraction pipeline fails. Table \ref{tab:multimodal_results} summarizes the results of our testing in terms of classification accuracy and F\textsubscript{1} score.

\begin{figure}[!t]  
    \centering
    \includegraphics[width=0.95\columnwidth]{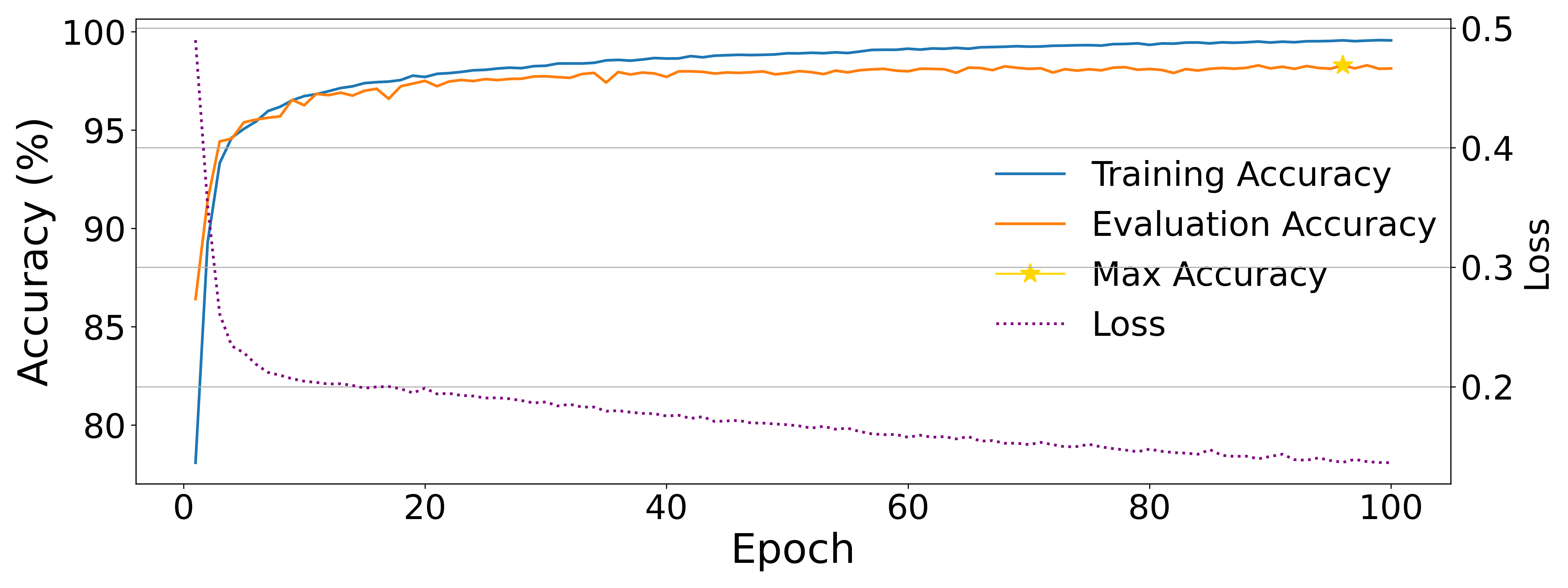}
    \vspace{-2mm}
    \caption{Malformer Training Accuracy with Loss}
    \label{fig:malformer_accuracy_plot}
\end{figure}

\begin{figure}[htbp]
\centering
\begin{subfigure}[b]{0.46\linewidth}
  \centering
  \includegraphics[width=\linewidth]{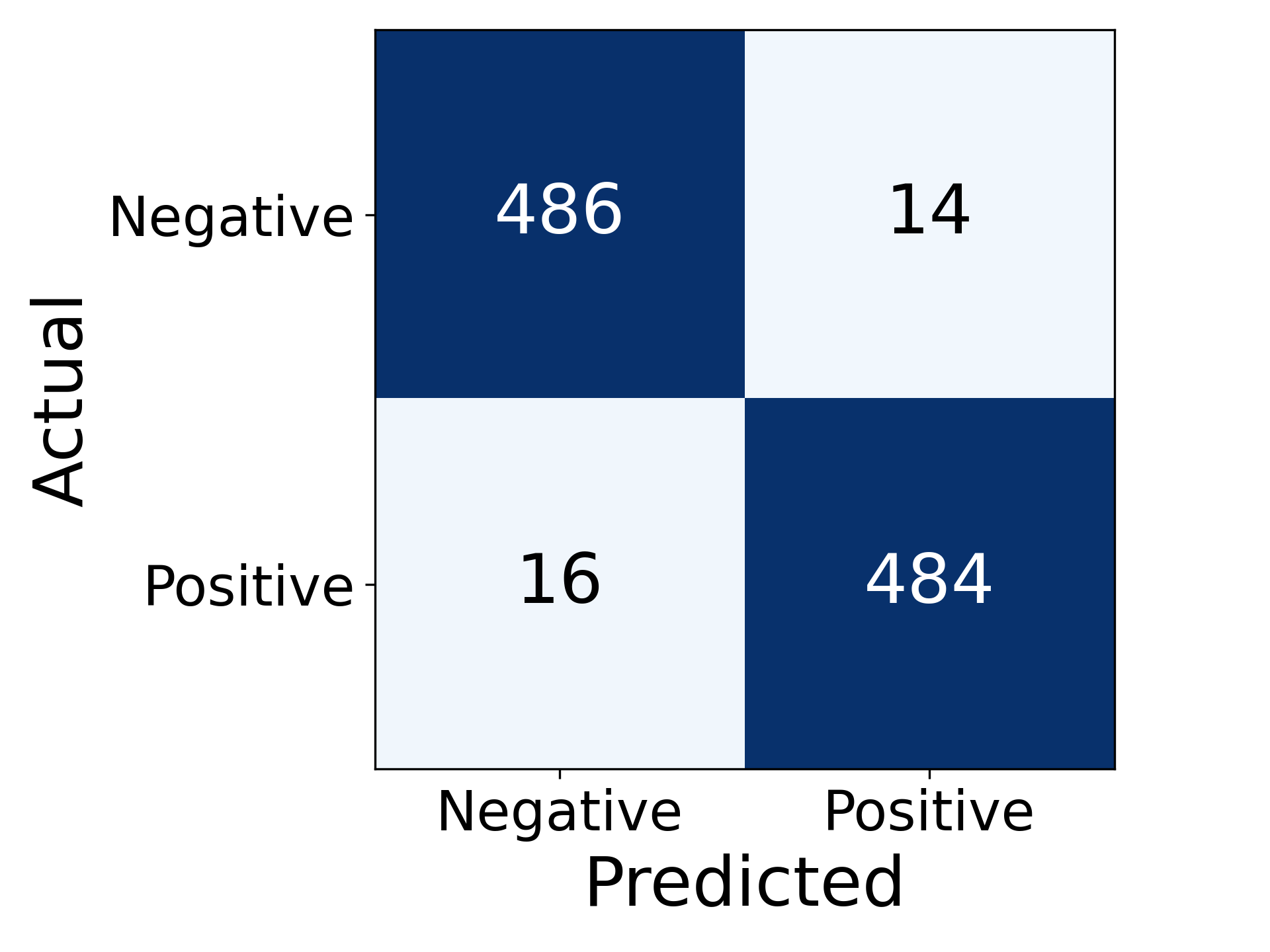}
  \caption{Small dataset}
\end{subfigure} \hfill
\begin{subfigure}[b]{0.46\linewidth}
  \centering
  \includegraphics[width=\linewidth]{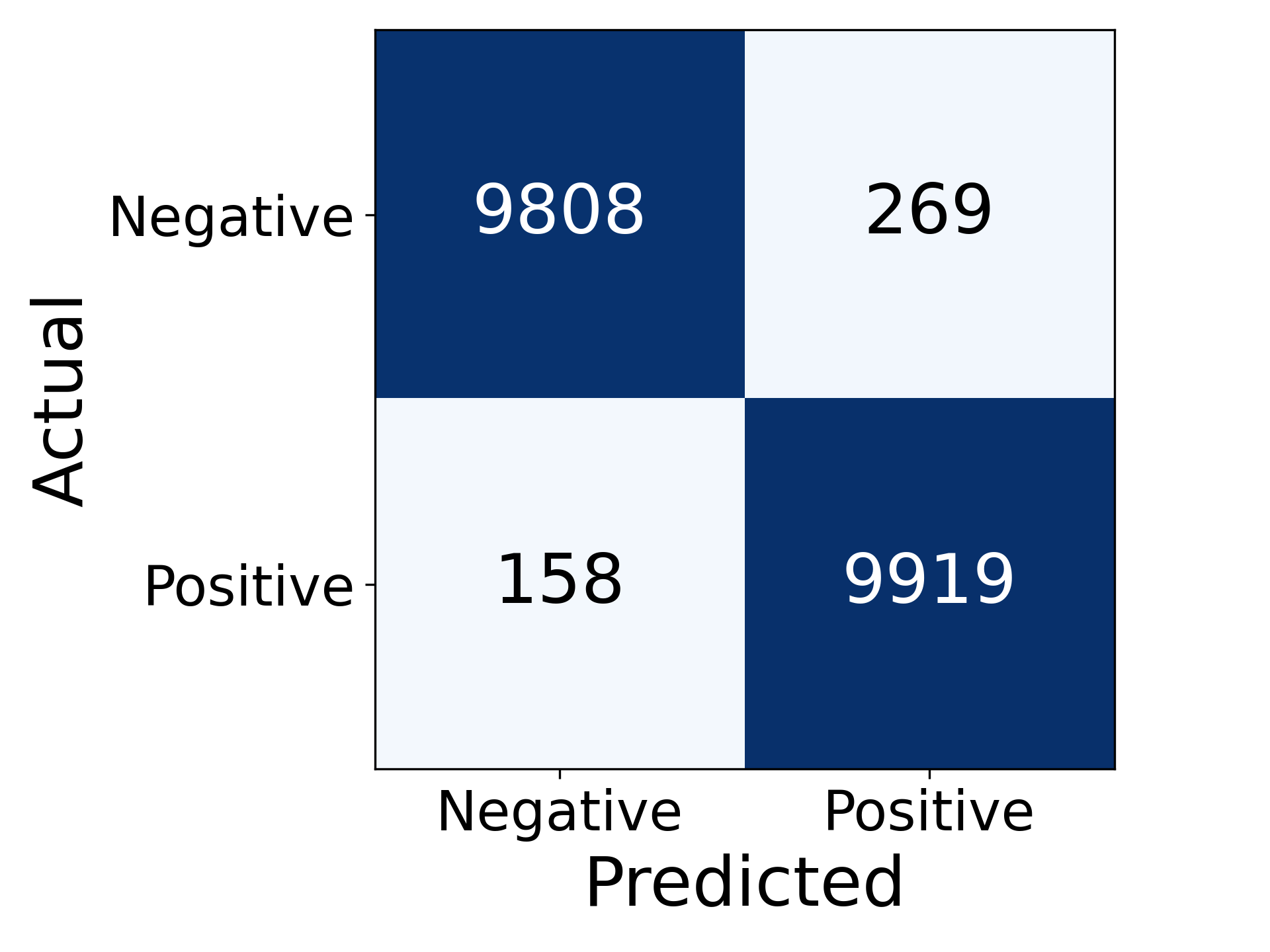}
  \caption{Large dataset}
\end{subfigure}
\caption{Confusion matrices for Malformer on both datasets}
\vspace{-3mm}
\label{fig:malformer_confusion_plots}
\end{figure}

In training, Malformer fits relatively rapidly, as shown in Figure \ref{fig:malformer_accuracy_plot}. Accuracy began to level off by approximately epoch 10 as the accuracy on the training set approached 100\%. Beyond this point, improvements on the validation set were marginal, suggesting that the model had effectively saturated its capacity to generalize under the given configuration. This behavior suggests that the effective rank of the RoBERTa-based subnetworks may be too high, allowing some amount of overfitting to the features of the training set. Despite this, Malformer's generalization performance remains high. This is further illustrated in Figure \ref{fig:malformer_confusion_plots} with Malformers low error rates.

Across multimodal comparisons, Malformer consistently outperformed the baseline unimodal approaches that the quadrimodal model is composed of. Compared to the graph modality alone, Malformer yields an improvement of 2.8 percentage points in accuracy on the small dataset and an improvement of 3.3 percentage points in accuracy on the large dataset. This demonstrates the strength of our multimodal approach and its improvement over a more restricted, unimodal approach. This performance gap underscores the complementary strengths of the three modalities. Through the combination of coarse linear features, fine 2D features, and fine linear features, the Malformer produces a stronger classifier than the sum of its unimodal parts. For security applications where accuracy is paramount, the tradeoff for time and memory complexity justifies itself through its improvement in classification performance.

\begin{figure}[!t]  
    \centering
    \includegraphics[width=0.95\columnwidth]{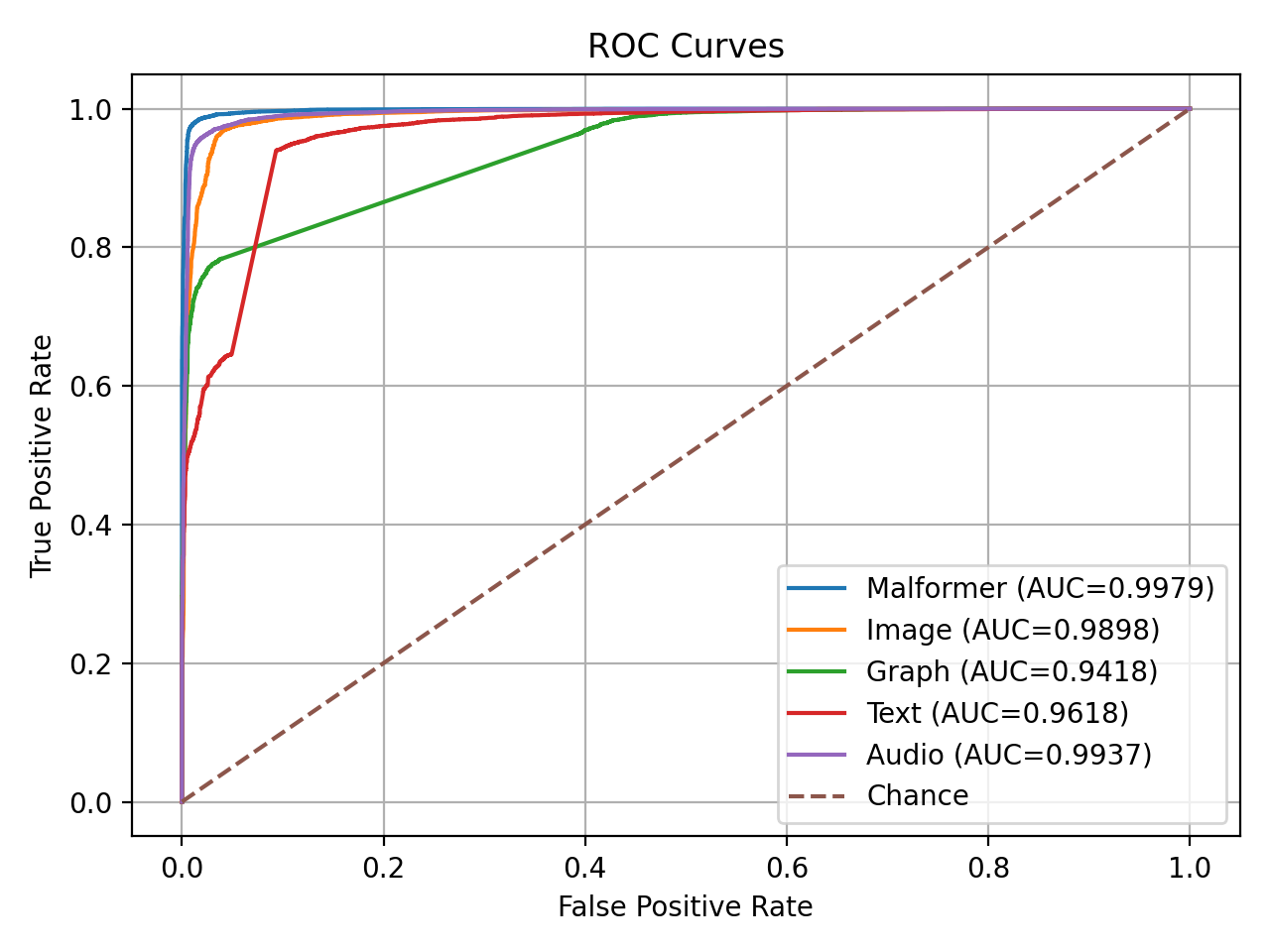}
    \vspace{-2mm}
    \caption{Malformer RoC curves for each decoder head}
    \label{fig:malformer_roc}
\end{figure}

Additionally, as shown in \ref{fig:malformer_roc}, Malformer shows excellent class separability overall, achieving an AUC of 0.9979. Of its modalities, image and audio perform similarly well to the overall model throughout, both dominating the text and graph modalities. The graph and text modalities have notably step-wise curves, indicative of their occasionally missing inputs. Still, the multimodal decoder head outperforms all the rest, demonstrating the benefits of our combination of orthogonal features for a better overall model.


Furthermore, as shown in Table~\ref{tab:multimodal_results}, Malformer outperformed both Orthrus and MIDALF by 5.1 and 5.0 percentage points in accuracy, respectively, on the Small dataset. On the Large dataset, Malformer extended its lead over MIDALF to 6.5 percentage points. A key factor in this improvement is the inclusion of the graph modality, which alone outperformed both Orthrus and MIDALF in our experiments. Orthrus relies on image and text, while MIDALF combines image with spectrogram-based audio features. By contrast, the graph modality encodes control-flow relationships that are more closely aligned with execution semantics, which likely contributes to its stronger standalone performance and its benefit within Malformer.

\begin{figure}[!t]  
    \centering
    \includegraphics[width=0.95\columnwidth]{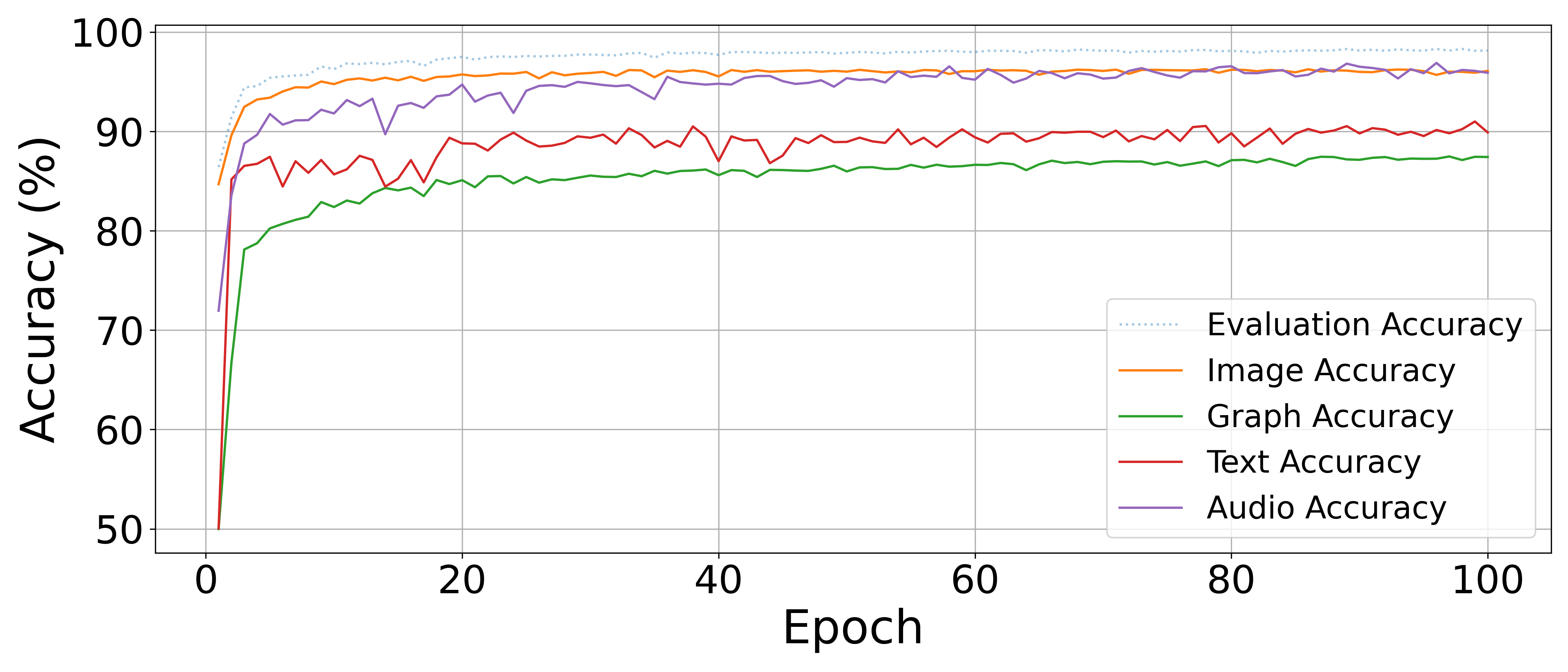}
    \vspace{-2mm}
    \caption{Malformer's Modality Training Accuracy by Epoch}
    \label{fig:modality_train_plot}
\end{figure}

Notably, some per-modality heads evaluated individually within Malformer performed marginally better or worse than their standalone counterparts, with Audio achieving 92.2\% accuracy and Text achieving 82.0\% accuracy on the small dataset, and Image achieving 96.0\% accuracy and Audio achieving 96.2\% accuracy on the large dataset. This suggests that the joint training process, combined with the adaptive loss-weighting schedule, may have provided mild regularization and calibration benefits compared to unimodal training (as shown in  Figure~\ref{fig:modality_train_plot}).

\subsubsection{Conditional Impact of Text and Graph Modalities}

In practice, assembly and control-flow graph extraction may fail due to obfuscation or disassembly errors, yielding a non-random pattern or missing modalities. To better understand this effect, we analyze Malformer’s performance conditioned on the successful extraction of the text and graph modalities. For this, we group evaluation samples from the large dataset according to whether graph features are present, text features are present, or both. Image and audio features are always available and are therefore omitted from the study. Table~\ref{tab:conditional_modalities} reports accuracy and F\textsubscript{1} score for each group.

When both graph and text features are present, performance is strong at 97.80\%, but slightly inferior to cases where graph features are present without text at 99.56\%. Conversely, when graph features are absent, text contributes meaningfully to classification performance at 99.28\%, but with an F\textsubscript{1} score of 0.9429 from an increase in false positives. This suggests that text features provide robustness against graph extraction failures but offer limited additional discriminative power when graph features are already available. This behavior aligns with the broader design goal of Malformer of robustness to partial feature failure, rather than guaranteed per-sample performance gains from all modalities. As such, the value of the text modality lies more in stabilizing performance across diverse extraction conditions than in enhancing classification accuracy when richer structural representations are available.

\begin{table*}[!t]
\centering
\caption{Performance based on availability of graph and text}
\vspace{-3mm}
\label{tab:conditional_modalities}
\begin{tabular}{lccc}
\hline
\textbf{Available Modalities} & \textbf{Samples} & \textbf{Accuracy} & \textbf{F$_1$ Score} \\
\hline
Image, Audio, Graph        & 1{,}428  & 99.56\% & 0.9956 \\
Image, Audio, Text         & 3{,}332  & 99.28\% & 0.9429 \\
Image, Audio, Graph, Text  & 13{,}423 & 97.80\% & 0.9787 \\
\hline
\end{tabular}
\end{table*}

\begin{table*}[!t]
\centering
\begin{threeparttable}
\caption{Model Complexity}
\begin{tabular}{lccc}
\hline
\textbf{Model} & \textbf{Modalities} & \textbf{GFLOPs} & \textbf{GPU Memory}\\
\hline
MIDALF & Image, Audio & 3.7 & 64.02 MB\\
Orthrus & Image, Text & 25.2 & 4572.00 MB\\
MalConv2 & Byte sequence & 139.1-543.5\tnote{a} & 53.69-145.54 MB \\
ViT & Image & 1.1 & 80.57 MB\\
WavLM & Audio & 6.9 & 496.30 MB\\
RoBERTa & Text or Graph & 173.9 & 544.05 MB\\
Malformer & \small{Image, Graph, Text, Audio}  & 362.7 & 1{,}664.00 MB\\  
\hline
\end{tabular}
\label{tab:model_flops_time}

\begin{tablenotes}
\footnotesize
\item[a] MalConv2 FLOPs and memory usage is dependent on input size. The given range is for input sequence lengths of 4,000,000 and 16,000,000. In our analysis, we used a sequence length of 16,000,000.
\end{tablenotes}
\end{threeparttable}
\end{table*}

\subsubsection{Model Complexity}

When comparing multimodal malware classifiers, it is important to consider both computational cost and memory usage. Table~\ref{tab:model_flops_time} summarizes the FLOP and GPU memory estimates for a single evaluation pass with one input tensor. Because FLOP counting was performed using different framework utilities for PyTorch- and TensorFlow-based models, the values should be interpreted as approximate indicators of relative complexity rather than exact cross-framework equivalence.


Although FLOP counts provide a useful baseline for comparing multimodal architectures, they do not fully reflect the practical computational constraints imposed by different design choices. MIDALF represents the lower end of the spectrum at 1.8 GFLOPs, relying on lightweight convolutional encoders for image and audio inputs that require minimal GPU memory and permit substantial batch sizes, but offer limited representational capacity. Orthrus, despite its moderate FLOP count of 12.6 GFLOPs, incurs a disproportionately high memory footprint of 4572.00 MB due to the large intermediate activations produced by its byte-level and opcode-level convolutional networks. On the Ubuntu server with an RTX 3060 (12GB of VRAM), it could not be trained with a batch size greater than one. In contrast, Malformer's transformer-based encoders yield the highest overall FLOP count of 362.7 GFLOPs, yet remain considerably more memory-efficient than Orthrus because their complexity arises primarily from attention operations rather than spatial activations. This difference enabled Malformer to train with batch sizes of approximately four on the same hardware, with the limiting factor becoming external bandwidth rather than GPU memory. Overall, compared to MIDALF and Orthrus, Malformer trades additional compute for stronger, semantically grounded representations while keeping its memory footprint manageable.

\begin{table*}[!t]
\centering
\caption{Ablation Study Results}
\vspace{-2mm}
\begin{tabular}{lccc}
\hline
\textbf{Missing Modality} & \textbf{Accuracy (\%)} & \textbf{Accuracy ($\Delta$)} & \textbf{F$_1$ Score} \\
\hline
image                  & 97.9          &  -0.4       & 0.9790 \\
graph                  & 95.0          &  -3.3       & 0.9524 \\
text                   & 96.4          &  -1.9       & 0.9654 \\
audio                  & 94.2          &  -4.1       & 0.9451 \\
image, graph           & 78.9          & -19.4       & 0.8255 \\
image, text            & 82.4          & -15.9       & 0.8500 \\
image, audio           & 92.4          &  -5.9       & 0.9259 \\
graph, text            & 93.1          &  -5.2       & 0.9351 \\
graph, audio           & 88.4          &  -9.9       & 0.8958 \\
text, audio            & 90.9          &  -7.4       & 0.9161 \\
\hline
\end{tabular}
\label{tab:ablation_results}
\vspace{-3mm}
\end{table*}

\subsection{Ablation Study}

For the ablation study, we evaluated the full quadrimodal model under single-modality and pairwise-modality removal. For the image and audio modalities, we replaced the encoder output with zeros. Next, for the RoBERTa-based modalities, we replaced the normal input with an encoded empty string, identical to the training process. This setup simulates feature-extraction failure for modalities that may not always be available in practice, particularly text and graph inputs.

Table \ref{tab:ablation_results} presents the results of our ablation study using the large dataset, where individual modalities were removed as described in Section \ref{malformer_method}. Most ablated versions outperform the bimodal baselines shown in Table \ref{tab:multimodal_results}, though results are worse with modalities removed, demonstrating the importance of each.

\subsubsection{Unimodal Ablation}

Removing the image modality is the least impactful of the four, with the model maintaining an accuracy of 97.9\%, a slight decrease of 0.4 percentage points from the quadrimodal result. This result, using text, graph, and audio, is significantly better than the image modality alone as a unimodal model. Removing just the audio or the graph modality produces similar results with decreases of 4.1 and 3.3 percentage points, respectively. Removing the text modality saw a modest decrease in accuracy of the trimodal configurations at 1.9 percentage points. Together, this demonstrates Malformer's resilience to attacks that cause the failure of single modalities in the pipeline, as the most reliable modalities together maintain consistent performance under the failure of other modalities.

Notably, when given a tensor composed completely of zeros or ones (and not zeroing the output), the image model heavily skews the prediction of the model.
This is potentially due to the method used to create the input images, padding shorter executables with zeros. This padding could be giving the model an unintended signal from the data based on the size of the executables. However, zeroing the output of the modality emulates the dropout regularization utilized during training, allowing the decoder to effectively ignore the image modality. Also of note, the image modality has an initial emphasis used to prevent overfitting of the RoBERTa-based text and graph modalities. This early dominance from our adaptive loss-weighting scheme may have led to a fused representation space for features from the RoBERTa-based modalities, leading to each complementing the other.

\subsubsection{Bimodal Ablation}


Removing pairs of modalities produces substantially larger performance drops than removing a single modality, confirming that the model benefits from combined information across modalities. The largest decreases occur when image is removed together with graph or text, with accuracy dropping to 78.9\% and 82.4\%, respectively.  With only image and text, there is an improvement over the baseline unimodal text model of 5.3 percentage points. Using only the image and graph modalities and omitting the text and audio data yields a 3.7 percentage point decrease in classification performance over the unimodal graph baseline. This indicates that both textual and graph-based inputs together provide essential signals that the model relies on for accurate predictions.

Combinations removing audio and either RoBERTa-based modality demonstrate a comparatively smaller decrease than that of image and either RoBERTa-based modality. Combined with the removal of audio, removing text results in a decrease in accuracy of 7.4 percentage points, and removing graph results in a decrease of 9.9 percentage points. This brings the bimodal part remaining down to below unimodal performance. However, removing the image modality and graph modality leads to a decrease of 19.4 percentage points, and removing the image modality and text modality leads to a decrease of 15.9 percentage points. These results demonstrate that the performance of Malformer's RoBERTa-based modalities is more aided by the representation generated by the image modality than that of the audio modality, while the RoBERTa-based models provide additional discriminative power to the full model when available.

The removal of both RoBERTa-based modalities or both image and audio yields significantly different results. As they are frequently missing due to feature extraction failure, removing both RoBERTa-based modalities only decreases accuracy by 5.2 percentage points, similar to removing the audio modality. Removing both image and audio decreases accuracy by 5.9 percentage points, demonstrating their combined importance to the efficacy of the model. These results demonstrate that within a multimodal system, consistency can be more important than feature information alone.

Overall, our results highlight that the value of a modality in our multimodal system is not solely determined by its unimodal accuracy, but also by its contribution to training dynamics and feature integration within the latent space of the model. The RoBERTa-based modalities form the backbone of the fused representation, while the image modality supplies complementary cues that improve generalization and robustness. Our results demonstrate not only the importance of each input and encoder network, but also the value of an adaptive fusion strategy to enable Malformer to exploit both synergistic and orthogonal relationships between modalities, rather than the strongest unimodal signal.

\subsubsection{Cross Analysis}

Given the relatively close performance of Malformer and MalConv2, we examined whether the two models make complementary errors and whether one model can improve decisions made by the other. To this end, we performed several cross-comparisons of the two models, attempting to see where one could aid the other the most and in what conditions. On the large dataset without packed samples, the two models performed somewhat close, with MalConv2 achieving 97.7\% accuracy and Malformer achieving 98.3\%. Between the two, both are wrong for only 0.67\% of samples. Additionally, MalConv2 correctly classifies samples Malformer gets incorrect for 1.04\% of the samples, and Malformer correctly classifies samples MalConv2 incorrectly classifies for 1.67\% of the samples. In total, this yields 2.70\% of samples for which the two models disagree, a potential gap for which a secondary model could fill. 

We initially started with the auxiliary decoder heads of Malformer for each modality, with image, graph, and text all decreasing the overall accuracy to $\sim$98.1\% when used to break disagreement ties between the two models. Audio, however, increased the overall accuracy to 98.46\%. This is a relatively marginal increase in accuracy when compared to the cost of adding MalConv2 to the equation, so another method would be preferred. We also trained XGBoost and LightGBM meta-classifiers on the prediction scores of the two models using disagreement samples from the training set. Because Malformer achieved higher training accuracy than MalConv2, these meta-classifiers were biased toward Malformer. On the evaluation set, XGBoost achieved 98.45\% accuracy and LightGBM achieved 98.41\%, which again indicates only limited improvement over Malformer alone.


\subsubsection{Unseen Samples}

\begin{table}[!t]
\centering
\caption{Evaluation on Unseen Samples}
\vspace{-2mm}
\begin{tabular}{lcc}
\hline
\textbf{Model} & \textbf{Accuracy (\%)} & \textbf{Recall (\%)} \\
\hline
Malformer      & 68.2                   & 88.2                 \\
MIDALF         & 57.0                   & 74.9                 \\
Orthrus        & 67.6                   & 70.2                 \\
MalConv2       & 51.6                   & 64.1                 \\
\hline
\end{tabular}
\label{tab:unseen_results}
\vspace{-3mm}
\end{table}

We further evaluated Malformer, MIDALF, Orthrus, and MalConv2 on our small dataset after being trained on the large dataset, as summarized in Table~\ref{tab:unseen_results}. We created a new subset of the small dataset by first filtering for valid PE files, then deduplicating against the large dataset by hash, yielding 18{,}872 samples, of which 12{,}583 were malicious. When evaluated on this new set, Malformer achieved an accuracy of 68.2\%, and MalConv2 achieved 51.6\%. MIDALF and Orthrus performed in the middle, at 57.0\% and 67.6\% respectively. Interestingly, Malformer's graph modality achieved the highest accuracy at 71.6\%.

The recall results show a different picture. MalConv2 achieves a recall of 64.1\%, Orthrus achieves 70.2\%, MIDALF achieves 74.9\%, and Malformer achieves 88.2\%. This indicates that, although overall accuracy drops substantially under distribution shift, Malformer retains a comparatively high malware detection rate. This discrepancy between overall accuracy and recall is largely explained by class imbalance and model bias. Although the unseen subset remained malware-heavy, several models exhibited a tendency toward predicting the benign class more frequently, resulting in reduced true positive rates despite moderate overall accuracy. MalConv2 in particular struggled to generalize to previously unseen malware, producing a large number of false negatives. In contrast, Malformer maintained a substantially higher detection rate, suggesting that its multimodal architecture learned more robust and transferable representations from the large dataset.

These results suggest that the multimodal architecture improves detection of malicious samples under distribution shift, but cross-dataset generalization remains challenging. Despite all overlapping binaries being removed through hash-based deduplication, Malformer retained a recall of 88.2\%, substantially exceeding the other multimodal baselines and MalConv2. This suggests that the additional modalities, particularly graph-based structural information, provided features that remained useful outside the original training corpus. At the same time, the comparatively modest gains in overall classification accuracy indicate that transferring between datasets remains challenging, likely reflecting differences in sample composition and benign software characteristics between the large and small datasets. Rather than uniformly improving all aspects of classification performance, training on the larger corpus primarily strengthened Malformer's ability to identify malicious samples under distribution shift.

\section{Conclusion}
\label{sec:conc}
In this work, we introduced \textit{Malformer}, a quadrimodal malware detection model that integrates image, text, graph, and audio modalities. Through the combination of these orthogonal perspectives of executable files, Malformer achieved an accuracy of 98.3\% and an F\textsubscript{1} score of 0.9833, surpassing both its unimodal baselines and existing bimodal approaches. Furthermore, ablation studies reveal the integral role of the more widely studied image modality as a foundational backbone for the fused representation space, with text, graph, and audio modalities providing further discriminative capability. Our findings demonstrate that effective malware detection arises not from strong unimodal models alone, but also integrated across a complementary feature space.

Malformer gives practitioners a generalized, robust defense against the diverse malware threats of the real world. Systems such as ours will only grow in importance as malware authors increasingly exploit vulnerable systems with greater volumes of novel malware variants. Multimodal detectors provide defenders with the tooling to oppose this trend automatically, at scale, and generalized across a plurality of attack vectors. 
An important future direction is testing the adversarial robustness of Malformer. Extending the application of Malformer to Android and Linux binaries would broaden its applicability across multiple operating systems.

\bibliographystyle{ACM-Reference-Format}
\bibliography{./bibtex/references}

\end{document}